\documentclass[]{emulateapj}

\usepackage{latexsym} 
\usepackage{amssymb}
\usepackage{amsmath}
\usepackage{graphicx}
\usepackage{tabularx}
\usepackage{booktabs}
\usepackage{hyperref}
\usepackage{fontawesome}
\usepackage{multirow}
\usepackage{xcolor}
\usepackage{adjustbox}
\usepackage{algorithm}
\usepackage{algpseudocode}
\usepackage{subcaption}

\graphicspath{{./}{./figures/}}
\begin{document}

\title{A Strong Gravitational Lens Is Worth a Thousand Dark Matter Halos: \\ Inference on Small-Scale Structure Using Sequential Methods\vspace{-4.5em}}
\shortauthors{Wagner-Carena et al.}
\shorttitle{A Strong Gravitational Lens Is Worth a Thousand Dark Matter Halos}

\author{{Sebastian~Wagner-Carena},\altaffilmark{1,2,6,*,\dag}
{Jaehoon~Lee},\altaffilmark{3}
{Jeffrey~Pennington},\altaffilmark{3}
{Jelle~Aalbers},\altaffilmark{4,6}
{Simon~Birrer},\altaffilmark{5}
{Risa~H.~Wechsler}\altaffilmark{6}}
\altaffiltext{1}{Center for Data Science, New York University} 
\altaffiltext{2}{Center for Computational Astrophysics, Flatiron Institute} 
\altaffiltext{3}{Google DeepMind} 
\altaffiltext{4}{Van Swinderen Institute for Particle Physics and Gravity, University of Groningen} 
\altaffiltext{5}{Department of Physics and Astronomy, Stony Brook University} 
\altaffiltext{6}{Kavli Institute for Particle Astrophysics and Cosmology, Department of Physics, Stanford University} 
\altaffiltext{*}{E-mail: \email{swagnercarena@simonsfoundation.org}}
\altaffiltext{\dag}{Work done in part while author was an intern at Google DeepMind}

\begin{abstract}

Strong gravitational lenses are a singular probe of the universe's small-scale structure --- they are sensitive to the gravitational effects of low-mass ($<10^{10} M_\odot$) halos even without a luminous counterpart. Recent strong-lensing analyses of dark matter structure rely on simulation-based inference (SBI). Modern SBI methods, which leverage neural networks as density estimators, have shown promise in extracting the halo-population signal. However, it is unclear whether the constraining power of these models has been limited by the methodology or the information content of the data. In this study, we introduce an accelerator-optimized simulation pipeline that can generate lens images with realistic subhalo populations in a matter of milliseconds. Leveraging this simulator, we identify the main methodological limitation of our fiducial SBI analysis: training set size. We then adopt a sequential neural posterior estimation (SNPE) approach, allowing us to iteratively refine the distribution of simulated training images to better align with the observed data. Using only one-fifth as many mock Hubble Space Telescope (HST) images, SNPE matches the constraints on the low-mass halo population produced by our best non-sequential model. Our experiments suggest that an over three order-of-magnitude increase in training set size and GPU hours would be required to achieve an equivalent result without sequential methods. While the full potential of the existing strong lens sample remains to be explored, the notable improvement in constraining power enabled by our sequential approach highlights that the current constraints are limited primarily by methodology and not the data itself. Moreover, our results emphasize the need to treat training set generation and model optimization as interconnected stages of any cosmological analysis using simulation-based inference techniques.

\end{abstract}

\section{Introduction}\label{sec:intro}\addtocounter{footnote}{-2}

The small-scale structure underlying our Universe plays a crucial role in distinguishing the fiducial cold, collisionless dark matter (CDM) model from alternative theories. Under CDM, the Universe's initial matter overdensities collapse into dark matter `halos'. These halos are predicted to form hierarchically \citep{white1978core, moore1999dark}, range in mass from cluster-scale ($10^{15} M_\odot$) to planetary masses \citep{navarro1996structure, navarro1997universal, green2004power}, and have an abundance that inversely scales with mass \citep{bode2001halo, kaplinghat2005dark, bullock2017small, buckley2018gravitational, tulin2018dark}. As a consequence, low-mass halos ($<10^{10} M_\odot$) should be found both as gravitationally isolated structures --- `line-of-sight' halos --- and as remnants within larger halos --- `subhalos'. Popular alternatives to CDM, including warm dark matter \citep{shi1999, viel2005, Schneider2012, lovell2014}, self-interacting dark matter \citep{Carlson1992, bullock2017small, kummer2018, vogelsberger2019}, and fuzzy dark matter \citep{hufidm2000, schive2016, hui2017}, significantly alter the formation history and properties of low-mass halos \citep{snowmass2021dm_from_halo}. Therefore, the abundance, distribution, and profiles of these halos directly constrain the physics of dark matter.

Among astrophysical probes of low-mass halos, galaxy--galaxy strong lensing is atypical in that it directly measures the gravitational influence of dark matter structure. A strong lensing image is created when the path of light from a distant `source' galaxy is curved by a `main deflector' galaxy to produce multiple images. In addition to the main deflector, the light's path is influenced by the line-of-sight halos it encounters along its path and the subhalos that reside in the main deflector. These additional perturbations produce a measurable signal that depends only on the mass distribution of the halos. Most alternative probes, including measurements of the Lyman-$\alpha$ forest (e.g.\ \citealt{rogers2021strong}), the UV galaxy luminosity function (e.g.\ \citealt{rudakovskyi2021constraints}), and the Milky Way's satellites (e.g.\ \citealt{nadler2021dark}), rely instead on a luminous tracer of the dark matter structure\footnote{Future precision measurements of gaps in stellar streams (e.g.\ \citealt{banik2021,
roman2024streamforecast}) also have the potential to detect dark halos, in a way that is complementary to strong lensing.}. Connecting low-mass dark matter structure to luminous tracers introduces significant uncertainties. By directly measuring the mass distribution, strong lensing provides an excellent complement to existing probes and has the potential to yield unprecedented insights into the nature of dark matter.

While the theory behind the strong lensing effect is well understood, inferring the low-mass halo population from images poses statistical and computational challenges. Most strong lensing problems aim to model only the light sources and the main deflector, allowing `traditional' Markov chain Monte-Carlo methods to be effective. However, the low-mass halo population can include thousands of line-of-sight halos and subhalos, each with parameters characterizing its position, mass, and profile. These hundreds of thousands of parameters will not be tightly constrained by the less than $20,000$ pixels in the image. There are two approaches to circumventing this limitation. `Direct' detection methods only search for the individual low-mass halos that significantly improve the model's fit to the data \citep{mao1998evidence, moustakas2003detecting, koopmans2005gravitational, vegetti2009bayesian, hezaveh2013dark}. These methods have been able to find subhalos in three strong lensing systems \citep{vegetti2010detection, vegetti2012gravitational, hezaveh2016detection, vegetti2018constraining, ccaugan2021substructure}. However, direct detection is insensitive to halos below $10^{9} M_\odot$ \citep{oriordan2023sensitivity}. While individual halos at these mass scales may not impart a statistically significant signal, the population can impart a collective signal that direct detection ignores.

``Statistical'' detection methods forgo constraining the exact distribution of low-mass halos and instead attempt to measure the properties of the entire population \citep{dalal2002direct, hezaveh2016measuring, cyr2016dark, birrer2017lensing, rivero2018gravitational, rivero2018power, brehmer2019mining, noemi2023wdm}. In fact, effects of alternative dark matter models are best understood in terms of statistical changes to the population, e.g.\ a depletion of halos below a certain mass scale \citep{viel2005, Schneider2012, lovell2014}, or a modification to the radial distribution of subhalos \citep{bullock2017small}). Most statistical modeling approaches focus on the abundance of subhalos as a function of mass, referred to as the subhalo mass function (SHMF). Modeling the SHMF introduces far fewer parameters and connects more directly to the physics of interest. However, framing the analysis in terms of the parameters of the SHMF makes evaluating the likelihood intractable. Even evaluating a single SHMF requires marginalizing over a broad range of configurations for thousands of subhalos. Without a likelihood, traditional modeling approaches are out of reach.

To overcome this challenge, statistical detection has focused on simulation-based inference (SBI) techniques\footnote{Historically, SBI has also been called likelihood-free inference.}. Broadly, SBI techniques leverage a simulator to implicitly define the likelihood. The classical method in the SBI family, approximate Bayesian computation (ABC) \citep{rubin1984bayesianly}, has already been used to constrain both alternative dark matter models \citep{gilman2020warm} and properties of the CDM model \citep{gilman2020constraints}. While ABC has shown success applied to data, it necessitates low-dimensional `summary' statistics that fundamentally limit the information that can be extracted. Recently, a new family of SBI methods has turned to using neural density estimators (see \citealt{sbi} for a review). These networks attempt to approximate either the likelihood (NLE) \citep{papamakarios2019sequential}, the posterior (NPE) \citep{lueckmann2017flexible}, or the likelihood ratio (NRE) \citep{mohamed2016learning}. In all cases, the resulting models take in the full data vector and are therefore not intrinsically limited by a compressed statistic. These methods are already prevalent in the strong lensing literature: they have been used to infer the parameters of the main deflector \citep{perreault2017uncertainties, wagner2021hierarchical, pearson2021strong} and the external convergence \citep{park2023kappa}. In the low-mass halo context, \cite{brehmer2019mining} used NRE to infer a Gaussian random field representing the subhalo population, \cite{noemi2023wdm} used NRE to constrain warm dark matter parameters, and \cite{zhang2024densityslope} used NRE to infer the density profile of subhalos. \cite{wagnercarena2022fidm} combined NPE with the most realistic simulation to-date to infer the SHMF normalization with up to one hundred lenses. Of these analyses, only \cite{zhang2024densityslope} has placed constraints on the data.

While these initial results are promising, they are still limited in their constraining power. For example, \cite{wagnercarena2022fidm} found only a weak correlation between the true and inferred SHMF parameters, and \cite{zhang2024densityslope} found a similarly weak correlation for the subhalo density slope. As these methods are pushed to the data, they will need to incorporate additional modeling and observational uncertainties that are likely to further degrade the constraints. Most of these analyses have also narrowed their focus to one or two population parameters, marginalizing over priors derived from simulations or external probes for the remaining parameters. Preferably, strong lensing would test these probes and simulations rather than depend on them. All neural density methods should extract the true, maximally constraining posterior, but only under idealized limits. This leads us to the central question we seek to answer: have neural-density-based SBI methods already reached the information limit of the data, or are there methodology choices imposing artificial bottlenecks?

In this paper, we identify the main methodological limitation behind the previous NPE analysis and leverage sequential neural posterior estimation (SNPE) to drastically improve the constraining power of the data. We build a new, accelerator-optimized simulation tool that allows us to generate batches of strong lensing images at the same rate as the optimization step of our neural network. Leveraging this `infinite' data limit, we explore four likely limitations to the results of \cite{wagnercarena2022fidm}: the complexity of the neural network model, the optimization strategy for gradient descent, the number of computing cycles dedicated to training the network, and the size of the training dataset. We focus our analysis on the normalization of the SHMF. Building on these results, we apply a sequential methodology (SNPE) that allows us to actively modify the proposal distribution from which training examples are drawn. We explore how steering the proposal distribution towards the observations affects model performance. Finally, we conduct a hierarchical analysis of a test population of thirty-two lenses and compare the constraints on the normalization of the SHMF for each method.

This paper is organized as follows. In Section \ref{sec:sim_methods} we discuss the parameterization we use to simulate our strong lensing systems and present our new, jax-optimized simulation package. Section \ref{sec:methods} introduces the simulation-based inference techniques we use throughout the paper, including neural posterior estimation (NPE), our hierarchical inference scheme, and the sequential extension to NPE (SNPE). Using our new simulator, we explore potential limitations of the NPE methods in Section \ref{sec:limitations}. Given the results of Section \ref{sec:limitations}, in Section \ref{sec:seq_inf_results} we implement the SNPE approach and contrast the resulting loss with that of NPE. To further compare the results of our sequential and non-sequential methods, in Section \ref{sec:hier_inf_results} we run both on a test population of thirty-two mock observations and extract the population-level constraints on the SHMF normalization. We discuss the limitations and future potential of our analysis in Section \ref{sec:discussion}. Section \ref{sec:conclusion} concludes by summarizing our results and contextualizing our findings within the broader scope of SBI applications for cosmological analysis.

\section{Simulation Methods}\label{sec:sim_methods}

The methodology tests we conduct in this paper require simulations that are representative of the data and can be generated in milliseconds. In the subsections that follow, we describe the parameterizations used to generate our lensing configuration and briefly outline \textsc{paltax} -- the new accelerator-optimized simulation tool we have built for this work. The parameterization of the lensing system we use in this work follows that of \cite{wagnercarena2022fidm} with minor modifications. Our lensing system can be broken into five components: the main deflector (Section \ref{sec:lp_main_deflector}), the subhalos of the main deflector (Section \ref{sec:lp_subhalos}), the source light (Section \ref{sec:lp_source}), the line-of-sight halos (Section \ref{sec:lp_los}), and the observational effects (Section \ref{sec:lp_obs_effects}). The details of our simulation pipeline, which is implemented using \textsc{jax}\footnote{\url{https://github.com/google/jax}} \citep{jax2018github}, are discussed in Section \ref{sec:jax_implementation}. All of the code used in this work has been thoroughly documented and tested.

\subsection{Main Deflector}\label{sec:lp_main_deflector}

The main deflector is modeled using a combination of power-law elliptical mass distribution (PEMD) and external shear. The PEMD profile \citep{kormann1994isothermal,barkana1998fast} is given by the convergence:
\begin{align}\label{eq:PEMD}
    \kappa(x,y) = \frac{3 - \gamma_\text{lens}}{2} \left( \frac{\theta_E}{\sqrt{q_\text{lens}x^2 + y^2 / q_\text{lens}}}\right)^{\gamma_\text{lens}-1},
\end{align}
where $\gamma_\text{lens}$ is the logarithmic power-law slope, $\theta_E$ is the Einstein radius, and $q_\text{lens}$ is the ellipticity. The $x$- and $y$-coordinates in Equation \ref{eq:PEMD} are defined along the major and minor axes of the PEMD profile. Therefore, there are three remaining degrees of freedom: the center of the PEMD profile, given by $x_\text{lens}$ and $y_\text{lens}$, and the orientation of the PEMD profile, given by $\phi_\text{lens}$. The external shear is given by a modulus $\gamma_\text{ext}$ and orientation angle $\phi_\text{ext}$ \citep{keeton1997shear}. The center of the external shear component is kept fixed at the center of the image.

We transform our main deflector parameterization to avoid inferring the two cyclical parameters, $\gamma_\text{ext}$ and $\gamma_\text{lens}$. We adopt an eccentricity/Cartesian coordinate system for our ellipticity/shear:
\begin{align}
    e_1 &= \frac{1-q_{\text{lens}}}{1+q_{\text{lens}}} \cos (2 \phi_\text{lens})\\
    e_2 &= \frac{1-q_{\text{lens}}}{1+q_{\text{lens}}} \sin (2 \phi_\text{lens})\\
    \gamma_1 &= \gamma_\text{ext} \cos (2 \phi_\text{ext}) \\
    \gamma_2 &= \gamma_\text{ext} \sin (2 \phi_\text{ext}).
\end{align}

For the remaining details on the main deflector, see \cite{wagnercarena2022fidm}.

\subsection{Subhalos}\label{sec:lp_subhalos}

The main deflector's subhalos follow the distribution introduced by \cite{gilman2020warm} along with the modifications from \cite{wagnercarena2022fidm}. The mass function of our subhalos is given by:
\begin{align}\label{eq:shmf}
    \frac{d^2 N_\text{sub}}{dA \ dm_\text{sub}} &= \Sigma_\text{sub} \frac{m_\text{sub}^{\gamma_\text{sub}}}{m_{\text{pivot,sub}}^{\gamma_\text{sub}+1}},
\end{align}
where $\Sigma_\text{sub}$ is the normalization of our SHMF, $\gamma_\text{sub}$ is the slope of our SHMF, $m_\text{sub}$ is our subhalo mass according to the $M_{200,c}$ mass definition \citep{white2001mass}, $dA$ is the differential area element, and $m_{\text{pivot,sub}}$ is the pivot mass. Subhalos are rendered between a minimum subhalo mass, $m_\text{min,sub}$, and a maximum subhalo mass, $m_\text{max,sub}$. For details on the mass--concentration relationship, radial distribution, and profile of the subhalos, see \cite{wagnercarena2022fidm}.

\begin{figure*}
    \centering
	\includegraphics[width=\textwidth]{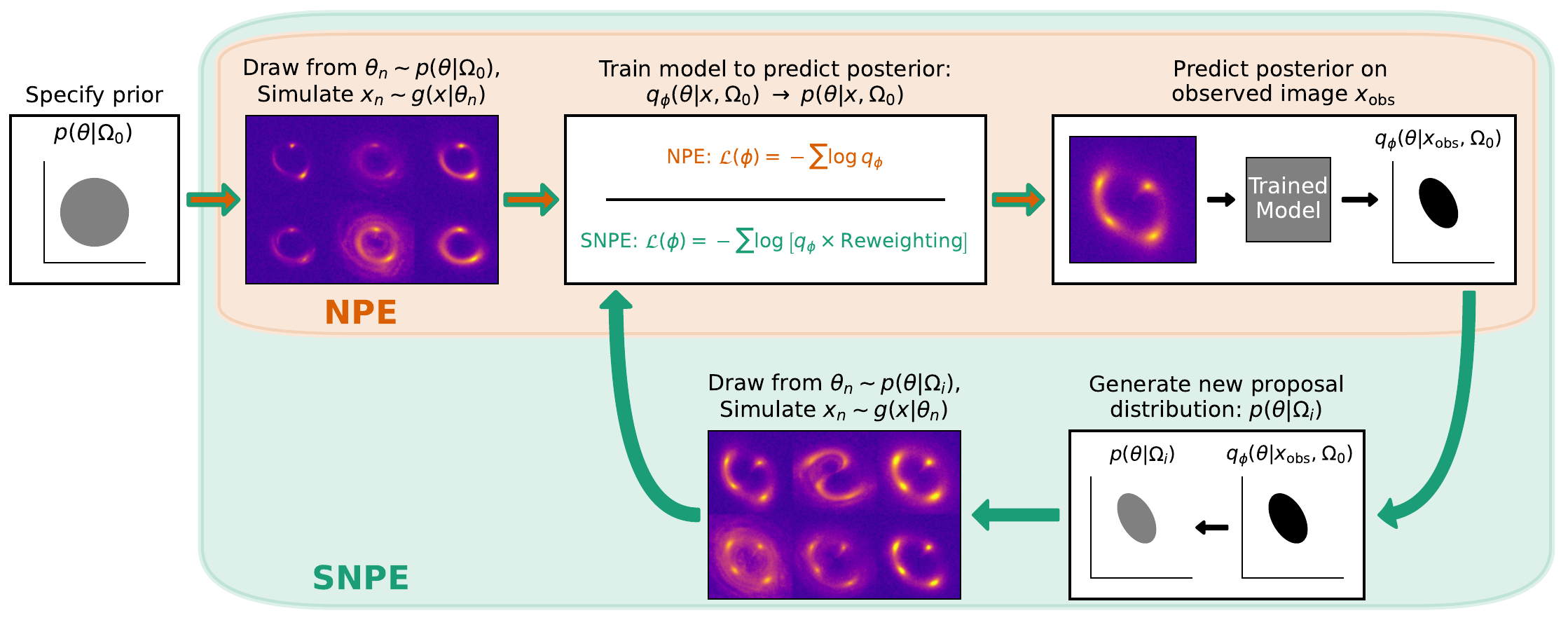}
    \caption{A schematic diagram comparing our two inference methods: neural posterior estimation (NPE, Section \ref{sec:meth_npe}) and sequential neural posterior estimation (SNPE, Section \ref{sec:meth_snpe}). Filled orange arrows with green outlines represent shared steps between NPE and SNPE. Green arrows are steps unique to SNPE. The $i$ index in the figure represents the $i^\text{th}$ iteration of sequential inference. In this figure, $g$ represents our stochastic simulator, $q_\phi$ is a conditional density estimator with parameters $\phi$, and $\mathcal{L}$ is the loss function. The distributions $p(\theta|\Omega)$ represent the prior on our parameters $\theta$ given the hyperparameters $\Omega$, and $p(\theta|x,\Omega)$ represents posterior on our parameters given an observation $x$. For more details, see Section \ref{sec:methods}.}
    \label{fig:method_comparison}
\end{figure*}

\subsection{Source}\label{sec:lp_source}

For the source light in our simulations, we use galaxy images drawn from the HST COSMOS survey \citep{koekemoer2007cosmos}. The source galaxies are selected from postage stamps used in the GREAT3 gravitational lensing challenge \citep{mandelbaum2012precision,mandelbaum2014third} with further cuts imposed on the minimum cutout size, apparent magnitude, half-light radius, and redshift. The cuts imposed are identical to those used in \cite{wagnercarena2022fidm}, resulting in a catalog of 2,262 source images\footnote{We refer the reader to \cite{wagnercarena2022fidm} for a more detailed discussion of the source selection cuts and the COSMOS survey.}. Of those galaxy images, 2,163 are used to generate training images, and 99 are reserved for generating validation images. Any step that involves updating the weights of the neural network (including for the sequential methods) uses the 2,163 training images. All of the metrics reported in this paper use the 99 validation images. 

The model for the source light is a linear interpolation of our pixelated galaxy images. Three parameters are introduced with the linear interpolation: the center of the source, given by $x_\text{source}$ and $\ y_\text{source}$, and the orientation of the galaxy, given by $\phi_\text{source}$. We keep the physical size of the galaxy fixed to the COSMOS observation, with the angular size rescaled to the simulated source redshift, $z_\text{source}$. Throughout this work, the source is assumed to be at a constant redshift of $z_\text{source} = 1.5$. For the amplitude, we scale the measured flux to $z_\text{source}$ and apply a k-correction. In lieu of a more complex modeling of the galaxy spectra, the k-correction is calculated assuming a flat spectral wavelength density. We also correct the offset in the zeropoint between the COSMOS observations and our target detector. The COSMOS galaxies were measured using the HST Advanced Camera for Surveys (ACS) with the F814W filter \citep{acs_inshandbook}. For our catalog, we assume a fixed AB zeropoint of 25.95, which is the average ACS zeropoint during the COSMOS survey \citep{koekemoer2007cosmos,mandelbaum2014third,acs_inshandbook}. Finally, we allow a multiplicative correction to the amplitude, given by $a_\text{source}$. 

\subsection{Line-Of-Sight Halos}\label{sec:lp_los}

In \cite{wagnercarena2022fidm}, the line-of-sight halos were included in the lensing simulations. Currently, the implementation of line-of-sight halos in \textsc{paltax} does not correct for the mean expected convergence. This would result in sightlines whose matter density is systematically higher than the universe's average \citep{birrer2017line}. For now, rather than introduce this bias to our simulations, we have elected to exclude the line-of-sight halo population. We hope to add this feature and thereby include line-of-sight halos in future work. Note that, in \cite{wagnercarena2022fidm}, it was shown that shifting the distribution of line-of-sight halos did not induce a detectable bias in the inferred subhalo mass function normalization.

\subsection{Observational Effects}\label{sec:lp_obs_effects}

Our simulations include observational effects to emulate the HST Wide Field Camera 3 (WFC3) UVIS channel with the F814W filter. This includes a pixel size of 0.40 arcsec/pixel (\citealt{wfc3_datahandbook}, section 1.1), a read noise of $3e^{-}$(\citealt{wfc3_datahandbook}, section 5.1.2), an AB magnitude zeropoint of 25.127 (\citealt{calamida2021new}), and a sky brightness\footnote{Calculated using \url{https://etc.stsci.edu/etc/input/wfc3uvis/imaging/}.} of 21.83 magnitude/arcsec\textsuperscript{2}. For the point spread function (PSF), we assume a simple Gaussian PSF with a full width at half maximum of $0.04 \arcsec$. Unlike in \cite{wagnercarena2022fidm}, we do not assume our images have gone through HST's drizzling pipeline, and therefore we keep the native pixel scale of $0.04 \arcsec$ for our final images.

\subsection{\textsc{paltax} Implementation}\label{sec:jax_implementation}

The main performance improvements of our simulation package over previous work come from the implementation of the full simulation pipeline in \textsc{jax}. This includes an implementation of a code for basic cosmological and large-scale-structure calculations, a ray-tracing code for the gravitational lensing calculations, a code that implements the observational effects from the telescope, and a code that samples the underlying lensing configuration to be simulated. The exact implementation details can be found in the github repository\footnote{\url{https://github.com/swagnercarena/paltax}}. The functionalities contained in \textsc{paltax} were previously divided among three packages: \textsc{paltas}\footnote{\url{https://github.com/swagnercarena/paltas}} \citep{wagnercarena2022fidm}, \textsc{lenstronomy}\footnote{\url{https://github.com/lenstronomy/lenstronomy}} \citep{birrer2018lenstronomy,birrer2021lenstronomy}, and \textsc{colossus}\footnote{\url{https://bitbucket.org/bdiemer/colossus}} \citep{diemer2018colossus}. With all the functionality under one package, we can leverage \textsc{jax} to seamlessly compile and vectorize the full simulation code. Note that the \textsc{paltax} repository also includes a robust test suite, including direct comparison to the results of \textsc{paltas}, \textsc{lenstronomy}, and \textsc{colossus}.

The timing improvement provided by our package is summarized in Table \ref{table:timing_comp}. In Appendix \ref{app:jax_implementation}, we outline some of the implementation choices required to achieve this performance. As a point of comparison, the fiducial network architecture we present in Section \ref{sec:limitations} takes 35 milliseconds\footnote{Timed on a NVIDIA Tesla V100 GPU} for a single training step on a batch of 32 images. Generating a batch of 32 images takes a little over 100 milliseconds with \textsc{paltax}, thereby enabling us to generate images on the fly as we train the model.

\begin{table}
   \centering
   \renewcommand{\arraystretch}{1.3}
   \begin{tabularx}{\columnwidth}{@{}  p{2.0cm} p{3.2cm} X @{}}
   \toprule
   \raggedright \textbf{Batch Size} & \textsc{paltax} \textbf{Timing} & \textsc{paltas} \textbf{Timing} \\
   \midrule
   
   1 & 9 milliseconds & 6 seconds \\

   2 & 13 milliseconds & 13 seconds\textsuperscript{\dag} \\

   8 & 31 milliseconds & 47 seconds\textsuperscript{\dag} \\

   32 & 103 milliseconds & 198 seconds\textsuperscript{\dag} \\
   
   \bottomrule
   \end{tabularx}
   \caption{Timing Comparison for \textsc{paltax} and \textsc{paltas}}
   \tablecomments{A comparison of the timing between \textsc{paltax} and \textsc{paltas}. Timing with \textsc{paltax} was calculated using a NVIDIA Tesla V100 GPU. Timing with \textsc{paltas} was calculated using a CPU with 16 cores. \\
   \dag: The \textsc{paltas} code is not parallelized, so all images are generated serially.}
   \label{table:timing_comp}
\end{table}

\section{Inference Methods}\label{sec:methods}

In this Section, we describe our two inference methods: neural posterior estimation (NPE, Section \ref{sec:meth_npe}) and sequential neural posterior estimation (SNPE, Section \ref{sec:meth_snpe}). A schematic summary is shown in Figure \ref{fig:method_comparison}. Section \ref{sec:meth_hi} presents the hierarchical inference framework that allows us to combine our constraints on multiple images.

\subsection{Neural Posterior Estimation}\label{sec:meth_npe}

In neural posterior estimation (NPE), the goal is to train a conditional density estimator, $q_{\phi}(\theta|x, \Omega)$, that approximates the parameter posterior $p(\theta|x, \Omega)$. Here, $\phi$ are the network parameters, $x$ is an observation, and $\theta$ are the physical parameters of interest. We have also been explicit about the hyperparameters, $\Omega$, which define the prior distribution on our physical parameters, $p(\theta|\Omega)$. To train our density estimator, we first draw samples $\theta_i$ from our prior distribution $p(\theta|\Omega_0)$. Note that $\Omega_0$ denotes the specific distribution we use to generate our training sample. These physical parameter values are pushed through our simulator, $g$, to generate data samples $x_i$. Each pair $(\theta_i, x_i)$ represents a sample from the posterior $p(\theta|x_i, \Omega_0)$. The conditional density estimator is then trained using the loss function:
\begin{align}\label{eq:npe_loss}
    \mathcal{L}(\phi) = - \sum_{n=1}^{N_\text{train}} \log q_{\phi}(\theta_n|x_n, \Omega_0),
\end{align}
where $N_\text{train}$ is the size of our training set\footnote{In practice, we train over shuffled batches of the training set.}. In the limit of $N_\text{train} \to \infty$, and given a sufficiently flexible variational distribution $q$, the loss in Equation \ref{eq:npe_loss} is minimized when $q_{\phi}(\theta|x, \Omega_0) \to p(\theta|x, \Omega_0)$ (see \citealt{papamakarios2016fast}). 

\subsection{Sequential Neural Posterior Estimation}\label{sec:meth_snpe}

The quality of the posteriors returned by NPE will depend on our choice of the prior distribution, $p(\theta|\Omega_0)$. Generally, if our training samples are dense near the parameter values of our observed lenses $\{\theta_\text{obs}\}$, our model will be more constraining. Since we do not know the parameter values of the lenses in our sky a priori, the only way to generate training samples near $\{\theta\}_\text{obs}$ is to choose a broad distribution for $p(\theta|\Omega_0)$ and draw a large sample size. As a consequence, the majority of the training examples will be far from our regions of interest. As we will explore more fully in Section \ref{sec:limitations_training_set}, this can hamper our model's ability to constrain our parameters of interest.

Sequential neural posterior estimation (SNPE) addresses this challenge by iteratively refining the distribution from which we draw training examples. At a high level, SNPE starts from the initial, broad distribution, $p(\theta|\Omega_0)$, trains a neural posterior estimator to output an approximate distribution, $q_\phi(\theta|x, \Omega_0)$, and then uses that approximate distribution on a single observed image, $x_\text{obs}$, as a new proposal distribution, $ p(\theta|\Omega_1) = q_\phi(\theta|x_\text{obs}, \Omega_0)$. The new proposal distribution can be used to train a new approximate distribution, $q_\phi(\theta|x, \Omega_1)$, and the process can be repeated $i$ times to generate proposal distribution $p(\theta|\Omega_i)$ and approximate distribution $q_\phi(\theta|x, \Omega_i)$.

Because the new proposal distributions are constrained by an observed image, these distributions generate more informative training samples. However, the distributions $q_\phi(\theta|x_\text{obs}, \Omega_i)$ have the proposal distributions, $p(\theta|\Omega_i)$, as a prior. The proposal distribution is not a proper prior because it is influenced by the observed data. Therefore, $q_\phi(\theta|x_\text{obs}, \Omega_i)$ cannot be treated as a proper posterior. To make this distinction clear, the distribution $q_\phi(\theta|x_\text{obs}, \Omega_i)$ is said to approximate the `proposal' posterior:
\begin{align}
    p(\theta| x, \Omega_i) &= p(\theta| x, \Omega_0) \frac{p(\theta | \Omega_i) p(x|\Omega_0)}{p(\theta | \Omega_0) p(x|\Omega_i)}.
\end{align}

Each SNPE implementation \citep{papamakarios2016fast, lueckmann2017flexible, greenberg2019automatic} circumvents this limitation differently. In this work, we will focus on the automatic posterior transformation \citep{greenberg2019automatic}, which we will refer to as SNPE-C. SNPE-C uses training examples drawn from the proposal distribution, $\theta_n \sim p(\theta_n|\Omega_i)$, but implements a re-weighted objective that allows the network to approximate the true posterior. The modified objective function is given by: 
\begin{align}\label{eq:SNPE-C}
    \mathcal{L}(\phi) = - \sum_{n=1}^{N_\text{train}} \log \left[ q_\phi(\theta_n|x_n, \Omega_0) \frac{p(\theta_n|\Omega_i)}{p(\theta_n|\Omega_0)} \frac{1}{Z(x_n, \phi)} \right].
\end{align}
Here $q_\phi(\theta_n|x_n, \Omega_0)$ is the approximate distribution being output by the network and $Z(x_n, \phi)$ is a normalizing constant given by:
\begin{align}
    Z(x_n, \phi) = \int q_\phi(\theta|x_n, \Omega_0) \frac{p(\theta|\Omega_i)}{p(\theta|\Omega_0)} \ d\theta.
\end{align}
Given a sufficiently expressive approximate distribution and in the limit of $N \to \infty$, minimizing Equation \ref{eq:SNPE-C} yields $q_\phi(\theta|x, \Omega_0) \to p(\theta | x, \Omega_0)$.

\begin{table*}
   \centering
   \renewcommand{\arraystretch}{1.3}
   \begin{tabularx}{\textwidth}{@{}  p{3.0cm} p{2.0cm} p{3.8cm} p{2.5cm} p{2.8cm} X @{}}
   \toprule
   \textbf{Model Name} & \textbf{Stage Size} & \textbf{Block Configuration} & \raggedright \textbf{Base Filters per Convolution} & \textbf{Total  Parameters} & \textbf{FLOPs/image} \\
   \midrule
   
   ResNet 18 Very Small & $[ 2, 2, 2, 2]$ & ResNet Block & 8 & 227,638 &  $5.0 \times 10^{5}$\\

   ResNet 18 Small & $[ 2, 2, 2, 2]$ & ResNet Block & 16 & 803,222 & $1.0 \times 10^{6}$ \\

   ResNet 18 & $[ 2, 2, 2, 2]$ & ResNet Block & 64 & 11,578,838 & $4.0 \times 10^{6}$ \\

   ResNet 34 & $[3, 4, 6, 3]$ & ResNet Block & 64 & 21,694,422 & $5.8 \times 10^{6}$\\

   ResNet 50 & $[3, 4, 6, 3]$ & Bottleneck ResNet Block & 64 & 23,953,878 & $1.6 \times 10^{7}$\\
   
   ResNet-D 50 & $[3, 4, 6, 3]$ & Bottleneck ResNet-D Block & 64 & 23,617,238 & $1.8 \times 10^{7}$\\
   
   \bottomrule
   \end{tabularx}
   \caption{Model Summary for Section \ref{sec:limitations_model_size}}
   \tablecomments{A comparison of the architectures used in Section \ref{sec:limitations_model_size}. A description of the ResNet architectures, along with a definition for the block configurations, can be found in \cite{he2016deep} and \cite{he2019bag}. Floating point operations (FLOPs) are per image and are estimated from the \textsc{XLA} compiled function generated by \textsc{jax}.}
   \label{table:model_size}
\end{table*}

SNPE-C allows for an informative proposal distribution while still yielding a model that outputs a proper posterior. In addition, because the prior is specified only in the objective function, it is possible to use a prior that would normally be impractical to sample from. As we will show in Section \ref{sec:seq_inf_results}, this can lead to significant improvements in the constraining power compared to NPE. 

However, there are a few limitations to the SNPE-C approach that are worth noting. First, the proposal distributions are tuned to specific observations. In fact, SNPE-C is traditionally trained with a single observation as the target. As a consequence, the model's improved performance will not be amortized, with each observation requiring its own model. It is possible to use a proposal distribution that mixes the posteriors from several observations, but the SNPE-C algorithm will still need to be re-run when new observations are acquired. Another shortcoming of the SNPE-C approach is that it can only refine the proposal distributions on parameters captured by the conditional density estimator, $q_\phi$. For example, the morphology of our sources is drawn from a set of COSMOS observations. Ideally, we would steer our proposal distributions to include more representative sources. However, since we do not predict any morphological parameters in the posterior, we cannot tune this aspect of the proposal distributions and must keep the initial sampling. Finally, to run SNPE-C we need the normalizing factor in Equation \ref{eq:SNPE-C}. For this work, we will use a multivariate Gaussian prior, posterior, and proposal distribution. In this limit, the normalizing factor is analytic. A non-analytic normalizing factor can also be dealt with using atomic proposals (see \citealt{greenberg2019automatic}), but we will not explore that regime here.

\subsection{Hierarchical Inference}\label{sec:meth_hi}

Both NPE and SNPE require us to enforce a desired prior $\Omega_0$ at training time. However, our final goal is to infer $\Omega_\text{obs}$, the hyperdistribution of parameters underlying our observed lenses $\{x_\text{obs}\}$. To conduct this inference, we place our (S)NPE outputs within a larger hierarchical model and calculate the posterior on $\Omega_\text{obs}$ given $\{x_\text{obs}\}$:
\begin{align}\label{eq:hier_inf_pos}
    p(\Omega_\text{obs} | \{ x_\text{obs}\})&= \!\begin{aligned}[t] & \underbrace{p(\Omega_\text{obs})}_\text{$\Omega_\text{obs}$ prior} \times \underbrace{\prod_k^{N_\text{lens,obs}} \frac{p(x_k|\Omega_0)}{p(\{x\})}}_\text{normalizing factor} \times \\
    & \underbrace{\prod_k^{N_\text{lens,obs}} \int \frac{p(\theta|\Omega_\text{obs})}{p(\theta|\Omega_0)} q_{\phi}(\theta|x_k, \Omega_0) \ d \theta}_\text{importance-sampling integral}, \end{aligned}
\end{align}
where $N_\text{lens,obs}$ is the number of observed lenses\footnote{See appendix C of \cite{wagner2021hierarchical} for a derivation of Equation \ref{eq:hier_inf_pos}.}. Equation \ref{eq:hier_inf_pos} re-weights our posterior to account for the difference between our interim proposal distribution, $\Omega_0$, and the distribution we are trying to infer on the observed lenses, $\Omega_\text{obs}$. This enables us to conduct inference on $\Omega_\text{obs}$ even though we cannot vary our choice of $\Omega_0$ within our posterior estimator, $q_{\phi}(\theta|x, \Omega_0)$.

We do not have an analytic solution for Equation \ref{eq:hier_inf_pos}. However, the three distributions involved in the integral, $p(\theta|\Omega_\text{obs})$, $p(\theta|\Omega_0)$, and $q_\phi(\theta|x_k,\Omega_0)$, will be Gaussians throughout this work. This gives us an analytic solution to the importance sampling integral in the equation, allowing us to use a sampling algorithm to generate representative draws from $p(\Omega_\text{obs} | \{ x_\text{obs}\})$. For this work, we use the ensemble sampler with affine invariance implemented in the \textsc{emcee} package\footnote{\url{https://emcee.readthedocs.io}} \citep{goodman2010ensemble, foreman-mackey2014}.

\begin{figure*}
    \centering
	\includegraphics[width=\textwidth]{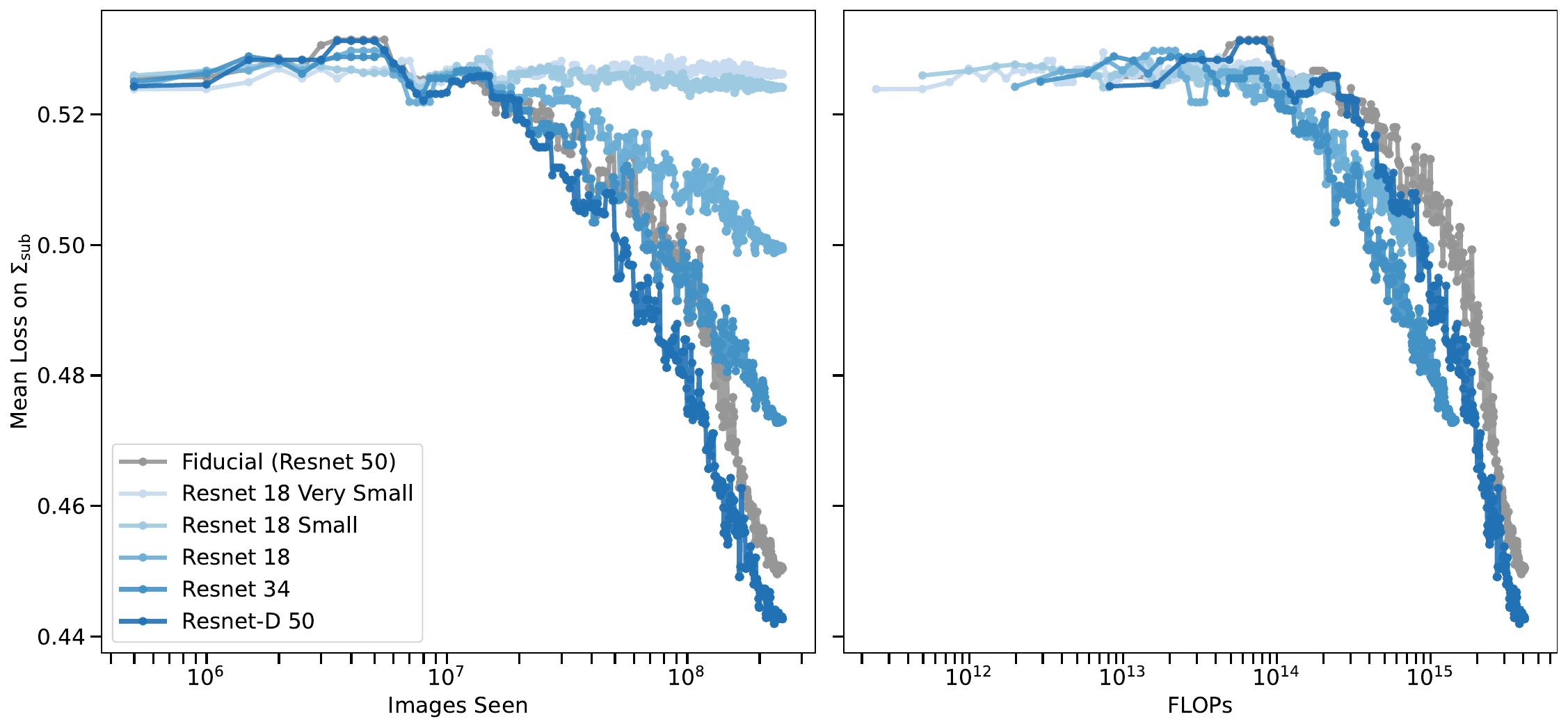}
    \caption{A comparison of the loss on $\Sigma_\text{sub}$ for five different model sizes: the fiducial ResNet 50 architecture (gray), a ResNet 18 architecture with a very small number of filters (lightest blue), a ResNet 18 architecture with a small number of filters (light blue), a traditional ResNet 18 architecture (blue), a ResNet 34 architecture (dark blue), and a ResNet-D 50 architecture (darkest blue). The left-hand plot shows the loss as a function of unique images seen, and the right-hand plot shows the loss as a function of the number of floating point operations.}
    \label{fig:lim_model_size}
\end{figure*}

\section{Results: Limitations on the Signal}\label{sec:limitations}

Our goal is to identify whether the NPE approach has already reached the information limit of our data or if there exist methodological choices that are diluting our constraining power. To distinguish between these two scenarios, we explore four possible methodological limitations:

\begin{itemize}
    \item \textbf{Model size} --- the NPE doesn't have enough parameters to capture the desired posterior. The small number of parameters is limiting the functional forms that our NPE can access during training.
    \item \textbf{Optimization strategy} --- a poor choice of learning rate schedule or optimizer is hindering the NPE's training. The NPE's weights are either converging to a local minimum, or the optimization strategy is substantially slowing convergence to the minimum.
    \item \textbf{Training set} --- our training set is too small / too sparse for the model to learn the optimal functional form. The broad training prior combined with the finite training set size means that the model does not have enough examples near the observed images.
    \item \textbf{Compute} --- there is no issue with the previous three choices. The model simply requires more iterations over the training set to better approximate the posterior.
\end{itemize}

To test these possibilities, we introduce a set of fiducial modeling choices. For this paper, we use a ResNet50 architecture \citep{he2016deep} trained using \textsc{Adam} as our optimization algorithm \citep{kingma2015adam}. For the learning rate, we use a cosine learning rate decay with a linear warmup \citep{loshchilov2017sgdr}. The learning rate has a maximum value of $0.01$, the linear warmup goes from $0.0$ to the maximum over 10 epochs, and the cosine decay occurs over 490 epochs. This results in a total of 500 epochs of training. For the fiducial configuration, each batch of training images fed to the model is unique and generated on the fly. The batch size is set to 32 images. An epoch of training is defined as 15600 steps, which gives roughly 500 thousand images per epoch\footnote{This is chosen to match the training set size of \cite{wagnercarena2022fidm}}.

The fiducial modeling choices are meant to improve on the choices made in \cite{wagnercarena2022fidm}. The model is larger, the optimization strategy conforms with current best practice, the training set size is effectively infinite, and the total number of training epochs is more than doubled. We will compare this fiducial set of choices to a number of test modeling choices. For each of our tests, we will specify how its choices differ from the fiducial set.

Throughout this work, our model will be predicting a multivariate Gaussian distribution with a diagonal covariance matrix. The model is trained to return the posterior for eleven lensing parameters. For the main deflector, this includes the Einstein radius, the power-law slope, the x- and y-coordinates for the center, the two ellipticity parameters, and the two shear parameters. For the source, this includes the x- and y-coordinates for the center. The final parameter is the subhalo mass function normalization, $\Sigma_\text{sub}$. The distribution of $\Sigma_\text{sub}$ values is well-predicted by CDM simulations \citep{nadler2021dark}, so a precise measurement of $\Sigma_\text{sub}$ would serve as a direct test of the fiducial dark matter model. It is both the parameter of interest and the parameter the model predicts most poorly, therefore our analysis will focus on the component of the loss associated with $\Sigma_\text{sub}$. The remaining ten parameters will play a larger role in Section \ref{sec:seq_inf_results}. The distribution of all of the simulation parameters can be found in Appendix \ref{app:sim_param_distributions}. All of the metrics presented in this section are calculated on 1024 test images. The lensing parameters of the test images are drawn from the same distribution used to train the model. The only exception is the source light. Because we have a finite number of COSMOS galaxy images to use as sources, 99 of our COSMOS images are excluded from training and used for our test images.

\begin{figure*}
    \centering
	\includegraphics[width=\textwidth]{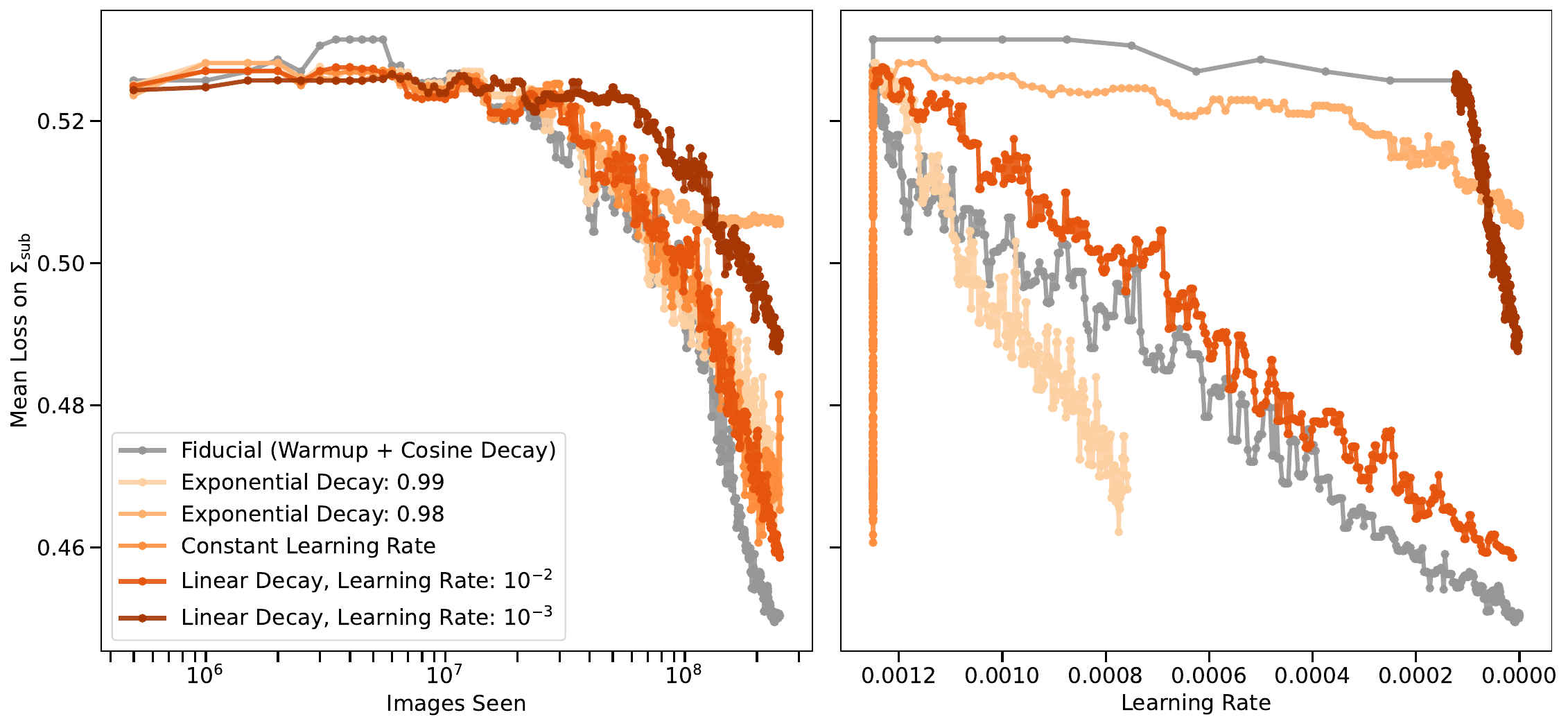}
    \caption{A comparison of the loss on $\Sigma_\text{sub}$ for six different learning rate schedules: the fiducial cosine learning rate decay with a warmup (gray), an exponential decay with a decay factor of 0.99 per epoch (lightest orange), an exponential decay with a decay factor of 0.98 per epoch (light orange), a constant learning rate (orange), a linear decay with a base learning rate of $0.01$ (dark orange), and a linear decay with a base learning rate of $0.001$ (darkest orange). The left-hand plot shows the loss as a function of unique images seen, and the right-hand plot shows the loss as a function of the learning rate. All of the learning rates are scaled by an additional 0.125 to account for the batch size of 32.}
    \label{fig:lim_learn_rate}
\end{figure*}

\subsection{Model Size}\label{sec:limitations_model_size}

The first limitation we explore is the size of the model. We compare six different model sizes: the ResNet-D 50 architecture \citep{he2019bag}, the fiducial ResNet 50 architecture, a ResNet 34 architecture, and three ResNet 18 architectures. The three ResNet 18 architectures differ in the number of filters being used per convolutional layer. Each model is detailed in Table \ref{table:model_size}. All other modeling choices conform to the fiducial configuration.

Figure \ref{fig:lim_model_size} shows the loss of the SHMF normalization for each of the six models, both as a function of the number of images seen (left-hand plot) and the total number of floating point operations (right-hand plot)\footnotetext{Throughout this work, loss curves are smoothed using a median filter with kernel size 5 to improve visualization.}\footnotemark[\value{footnote}]\newcounter{footcurve}\setcounter{footcurve}{\value{footnote}}. As a function of images seen (or equivalently the number of training steps), the size of the model correlates directly with the performance. The two largest models, the ResNet-D 50 and ResNet 50 architectures, achieve the lowest loss while the two smallest models, ResNet 18 Small and Very Small, never outperform their initial loss. However, as the size and complexity of a model increase, the number of floating point operations (FLOPs) required to conduct a training step also increases. FLOPs directly correlate with the resource cost of training the network and are therefore a better measure of how different architectures perform given a fixed computational budget\footnote{Generating a batch of image on-the-fly has a comparable computational cost to a step of gradient descent for the ResNet 50 architecture. Therefore, when the infinite training set is being used, the resource benefits of smaller models are less pronounced.}. When comparing the model performance as a function of FLOPs, model size no longer appears influential. All six models trace a similar learning curve, with the two smallest models --- ResNet 18 Very Small and ResNet 18 Small --- never fully crossing the apparent FLOPs threshold required to constrain $\Sigma_\text{sub}$. The ResNet 18 and 34 models, which do cross this threshold, closely follow the performance of the two ResNet 50 models.

Taken together, the results presented in Figure \ref{fig:lim_model_size} suggest that the architecture is not the dominant limitation in constraining $\Sigma_\text{sub}$. In fact, there is no evidence that the ResNet 18, ResNet 34, ResNet 50, or ResNet50-D architectures reach the limits of their constraining power. Instead, all four models continue to improve their loss when they are given more computational budget and more unique images. This does not mean that model size is irrelevant for strong lensing analysis, only that it is not the modeling choice currently limiting our ability to constrain low-mass halos.

\subsection{Optimization Strategy}\label{sec:limitations_optimization}

\begin{figure*}
    \centering
    \includegraphics[width=\textwidth]{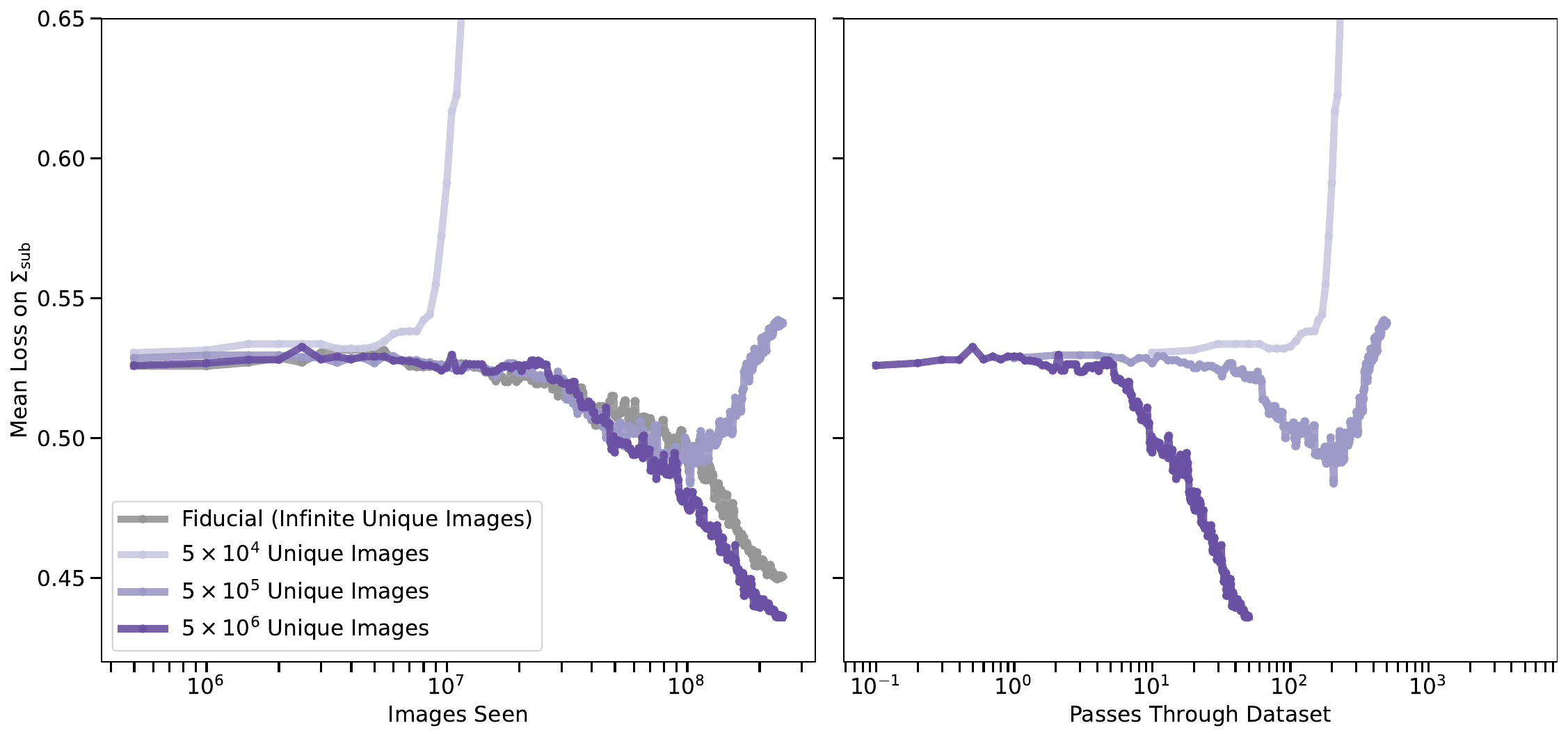}
    \caption{A comparison of the loss on $\Sigma_\text{sub}$ for four different training set sizes: the fiducial set with effectively infinite images (gray), a training set with 50 thousand unique images (light purple), a training set with 500 thousand unique images (purple), and a training set with 5 million unique images (dark purple). The left-hand plot shows the loss as a function of unique images seen, and the right-hand plot shows the loss as a function of the number of passes through the training set. Note that the fiducial choice (gray) is not shown on the right-hand side since the training dataset is effectively infinite.}
    \label{fig:lim_train_set}
\end{figure*}

Without evidence that model size limits our current performance, we move to testing the optimization schedule. Our fiducial model uses a cosine learning rate decay with a linear warmup. This schedule is defined by:
\begin{align}\label{eq:cosine_lr}
    l(s) &= \begin{cases}
        l_\text{base} \left( \frac{s}{s_\text{warm}} \right) & \text{if } s \leq s_\text{lin}\\
         \frac{l_\text{base}}{2} \left( 1 + \cos \left(\frac{s - s_\text{warm}}{s_\text{tot} -  s_\text{warm}}  \pi \right) \right) & \text{if } s > s_\text{warm},
    \end{cases} 
\end{align}
where $l(s)$ is the current learning rate at step $s$, $l_\text{base}$ is the base learning rate, $s$ is the current step, $s_\text{warm}$ is the number of warmup steps, and $s_\text{tot}$ is the total number of steps. The number of warmup steps is set to 10 epochs, the total number of steps is set to 500 epochs, and the base learning rate is set to 0.01.

We compare three alternate learning rate schedules: exponential decay, linear decay, and a constant learning rate. The exponential decay schedule is defined by:
\begin{align}\label{eq:exp_lr}
    l = l_\text{base} \left(d_\text{exp}\right)^{s / s_\text{epoch}},
\end{align}
where, $d_\text{exp}$ is the exponential decay rate and $s_\text{epoch}$ is the number of steps in an epoch. We test two different decay rates: $d_\text{exp}=0.99, \ 0.98$. The linear decay schedule decays from the initial base value, $l_\text{base}$ to a final learning rate $l_\text{final}$. The final learning rate is set to one one-hundredth that base rate, and we compare two different base rates $l_\text{base} = 0.01, \ 0.001$. For the constant learning rate, we keep the learning rate at the base value of $0.01$ throughout training.

In Figure \ref{fig:lim_learn_rate} we show the loss on the SHMF normalization as a function of images seen (left-hand plot) and the learning rate (right-hand plot). While the fiducial learning rate schedule performs the best, the constant schedule, the exponential decay schedule with a decay factor $0.01$, and the linear decay with base learning rate $10^{-2}$ perform nearly as well as our fiducial choice. Between our learning rates, we span a wide range of optimization schedules, yet with the exception of the exponential decay with a decay factor of $0.99$, all of our learning rates produce decreasing loss with an increasing number of images seen. There is also no clear correlation between learning rate and loss -- the constant schedule follows an almost identical trajectory in loss as the best-performing exponential schedule despite a several order-of-magnitude difference in the final learning rate. As with the model size, this does not mean that the choice of optimization schedule is irrelevant to our analysis. The fiducial choice is the best performer, and a sufficiently small learning rate can hamper performance. However, there appears to be a wide range of learning rate schedules that can successfully traverse the loss landscape. The main determinant of performance remains the number of images seen or, equivalently, the number of computations performed.

\subsection{Training Set Size and Compute}\label{sec:limitations_training_set}

\begin{figure*}
    \centering
	\includegraphics[width=\textwidth]{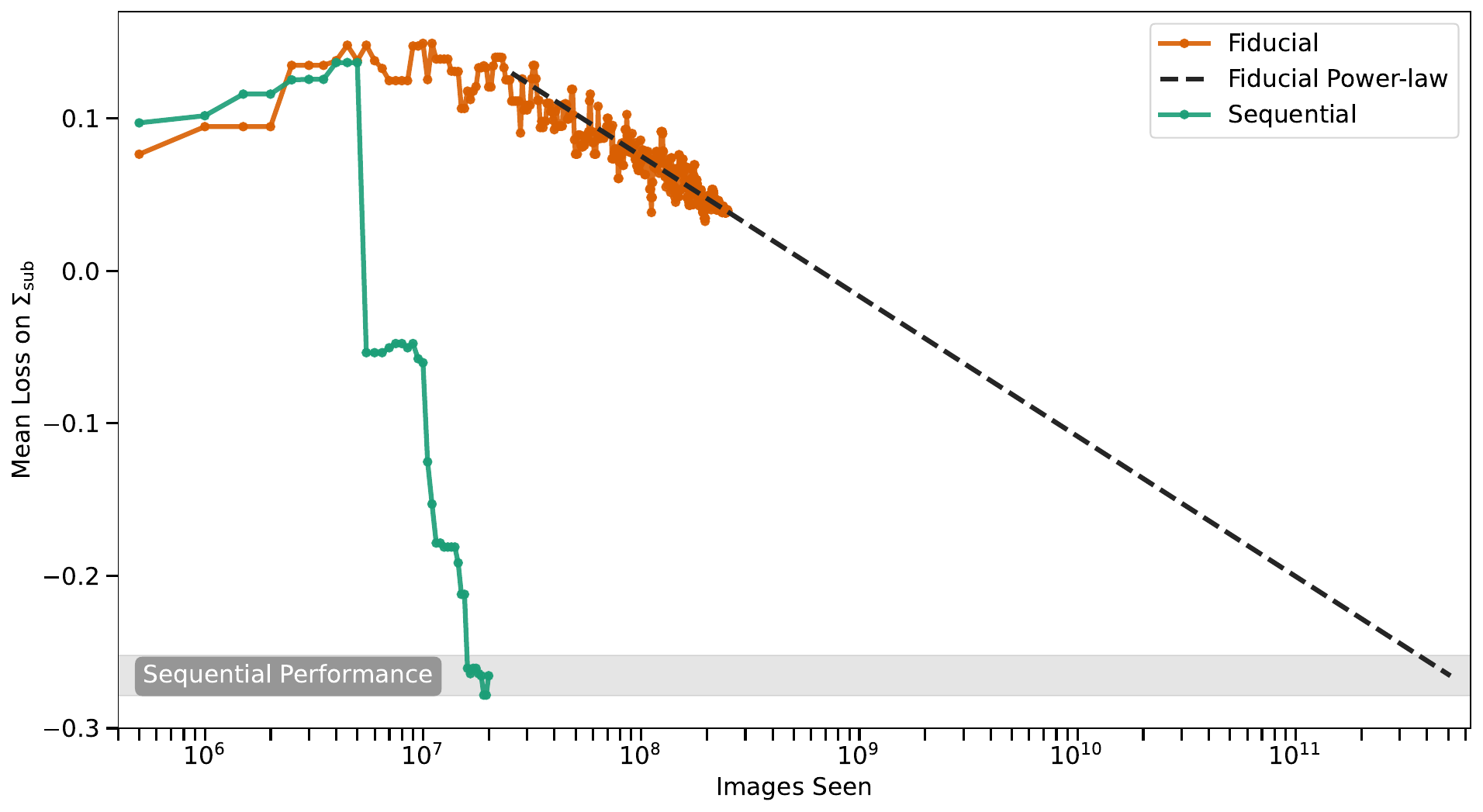}
    \caption{A comparison of the mean loss on $\Sigma_\text{sub}$ on 30 mock observations for two different methodologies: the fiducial, NPE approach (orange) and the sequential, SNPE approach (green). The dashed black line shows a power-law fit to the fiducial results using the loss for $> 2 \times 10^{7}$ images seen. The gray region bounds the sequential loss produced by the final proposal distribution. The jumps in sequential performance correspond to the transition between proposal distributions.}
    \label{fig:sequential_loss_sigma_sub}
\end{figure*}

In Section \ref{sec:limitations_model_size} and Section \ref{sec:limitations_optimization} we found that for a wide range of architectures and optimization schedules, the number of computations performed during training remained the main determinant of performance. However, our model sees a unique set of images with each batch. Therefore, we still need to disentangle the advantages of an increased computational budget from those of a larger training set. To test this, we artificially gate the number of unique images generated by our simulation pipeline during training. We also add random rotation augmentations to better reproduce the training choices that would be made with a finite dataset. We test three different dataset sizes: $5 \times 10^{4}$ unique images, $5 \times 10^{5}$ unique images, and $5 \times 10^{6}$ unique images. The dataset with size $5 \times 10^{5}$ images corresponds to the training set size used in \cite{wagnercarena2022fidm}.

Figure \ref{fig:lim_train_set} shows the resulting loss as a function of images seen (left-hand plot) and the number of passes through the dataset (right-hand plot). The training run with $5 \times 10^{6}$ unique images performs nearly identically to the fiducial, infinite training set size, but the two smaller training sets perform substantially worse. The right-hand plot suggests that after $\sim 100$ passes through the dataset, the ResNet 50 architecture begins to overfit to the training set. On the surface level, these results are not surprising: it is common for large architecture to overfit to small datasets, and a model with 24 million parameters should have the capacity to overfit to 50 thousand examples. However, it also demonstrates that the model's performance will not generically scale with the computational budget. The training set size must be scaled in tandem. For the $5 \times 10^{5}$ training set size used in \citep{wagnercarena2022fidm}, the ResNet 50 architecture is limited by the training set and not the computational budget.

Unlike the optimization schedule and architecture, our training set size is a strong determinant of the model's performance. When we are not limited by our training set, there is no evidence that our ResNet 50 model has saturated the information it can extract from our strong lensing data. In theory, our new simulation pipeline allows us to access an infinite training set and continue to move down the loss curve. However, the improvements in loss scale logarithmically with the number of images seen. Substantially improving the performance of our fiducial model would require an order of magnitude more computing resources. 

\section{Results: Sequential Inference}\label{sec:seq_inf_results}

The results of Section \ref{sec:limitations} suggest that our primary modeling limitation is our training set. In this section, we leverage SNPE to steer our proposal distribution towards more informative samples with the hopes of improving on the scaling seen in Figure \ref{fig:lim_train_set}. To test the SNPE approach, we create a new set of 30 test images. To distinguish these from the images used in Section \ref{sec:limitations}, we will label these our mock observations. The lensing parameters for the mock observations are drawn from the same distribution as our training set, with two exceptions. As with our previous test set, the source images are drawn from a held-out set of 99 COSMOS images. The distribution of SHMF normalizations has also been changed to approximately match a suite of high-resolution dark-matter-only (DMO) simulations \citep{nadler2013zoom}. This results in a shift in the mean normalization, $\Sigma_\text{sub,pop}$ from $2 \times 10^{-3} \text{kpc}^{-2} \to 1.5 \times 10^{-3} \text{kpc}^{-2}$ and a change in the population scatter, $\Sigma_{\text{sub,pop},\sigma}$, from $1.1 \times 10^{-3} \text{kpc}^{-2} \to 2 \times 10^{-4} \text{kpc}^{-2}$. This choice was made to quantify our constraining power on lenses that match our current theoretical expectations.

We run our sequential method on each of our 30 mock observations. For each sequential run, our proposal distribution is updated every 10 epochs of training (156000 steps or roughly 5 million images). This means that the fiducial, broad proposal distribution, $p(\theta|\Omega_0)$, is only used for the first 10 epochs of training. We run a total of 40 epochs of sequential training, meaning that each sequential model has access to four proposal distributions. The sequential training is stopped after 40 epochs because the models begin to overfit to the finite number of COSMOS sources being used (see Appendix \ref{app:seq_train_overfit} for details). The sequential training uses the same ResNet 50 architecture as the fiducial model, but the cosine decay is initialized to a base learning rate of $0.001$ rather than $0.01$. Finally, each new proposal distribution is set to be equal to the current estimate from the model:
\begin{align}
    p_{j}(\theta | \Omega_i) = q_{\phi,i-1}(\theta|x_j, \Omega_{0}),
\end{align}
where the $j$ index corresponds to the specific observation being considered by the sequential model and the $i$ index corresponds to the proposal / posterior from $i^\text{th}$ round of sequential inference. For a discussion of other proposal choices, see Appendix \ref{app:seq_proposal}.

\begin{figure}
    \centering
    \includegraphics[width=0.47\textwidth]{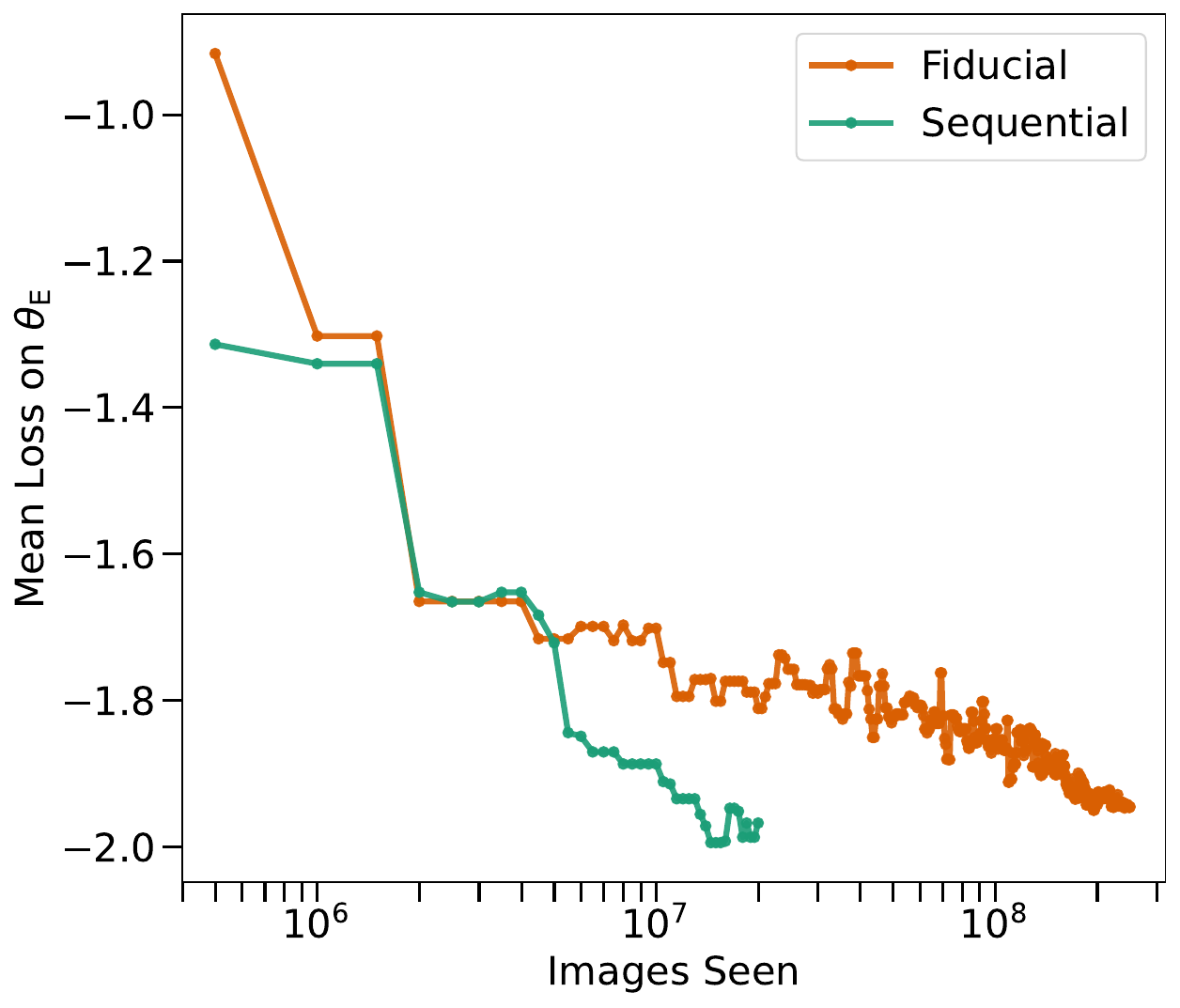}
    \caption{A comparison of the mean loss on $\theta_E$ on 30 mock observations for two different methodologies: the fiducial, NPE approach (orange) and the sequential, SNPE approach (green).}
    \label{fig:sequential_loss_theta_e}
\end{figure}

The mean loss on $\Sigma_\text{sub}$ on our 30 mock observations is shown in Figure \ref{fig:sequential_loss_sigma_sub}. The loss curve for the fiducial model follows a similar behavior to the larger test set used in Section \ref{sec:limitations}. By construction, the sequential models follow the same loss curve for the first 10 epochs. However, after a single epoch of training on the sequential proposal (the 11\textsuperscript{th} total epoch of training), the average sequential loss is already better than 500 epochs of training for the fiducial approach. Each additional sequential proposal leads to another jump in performance. By the end of the 40 epochs of training, the sequential approach drastically outperforms the fiducial approach. Figure \ref{fig:sequential_loss_sigma_sub} also includes a power-law fit to the fiducial loss curve for  $>2 \times 10^{7}$ images seen. If the fiducial loss continues along this power-law relationship, the NPE model would need to see $5.2 \times 10^{11}$ unique images to match the sequential results. That is equivalent to training the fiducial model 2,000 times longer.

Because the sequential approach requires one model per observed image, we can also compare how many observations we would need for the computational cost of both approaches to be equivalent. For example, for the thirty mock observations we analyzed, the 500 epochs of fiducial training have the same computational cost as 26 epochs of sequential training. The power law in Figure \ref{fig:sequential_loss_sigma_sub} suggests the fiducial model would require 1 million epochs of training to match the sequential approach. Under that estimate, we would need to analyze $26,000$ lenses for the fiducial and sequential approaches to have the same computational cost. Note that extrapolating along the power law in the fiducial approach's loss is a conservative estimate of the benefits of the sequential approach. Scaling four order-of-magnitude would likely reach the capacity limits of the ResNet 50 architecture, and the added computational cost of a larger architecture would further disadvantage the fiducial approach.

\begin{figure}
    \centering
    \includegraphics[width=0.47\textwidth]{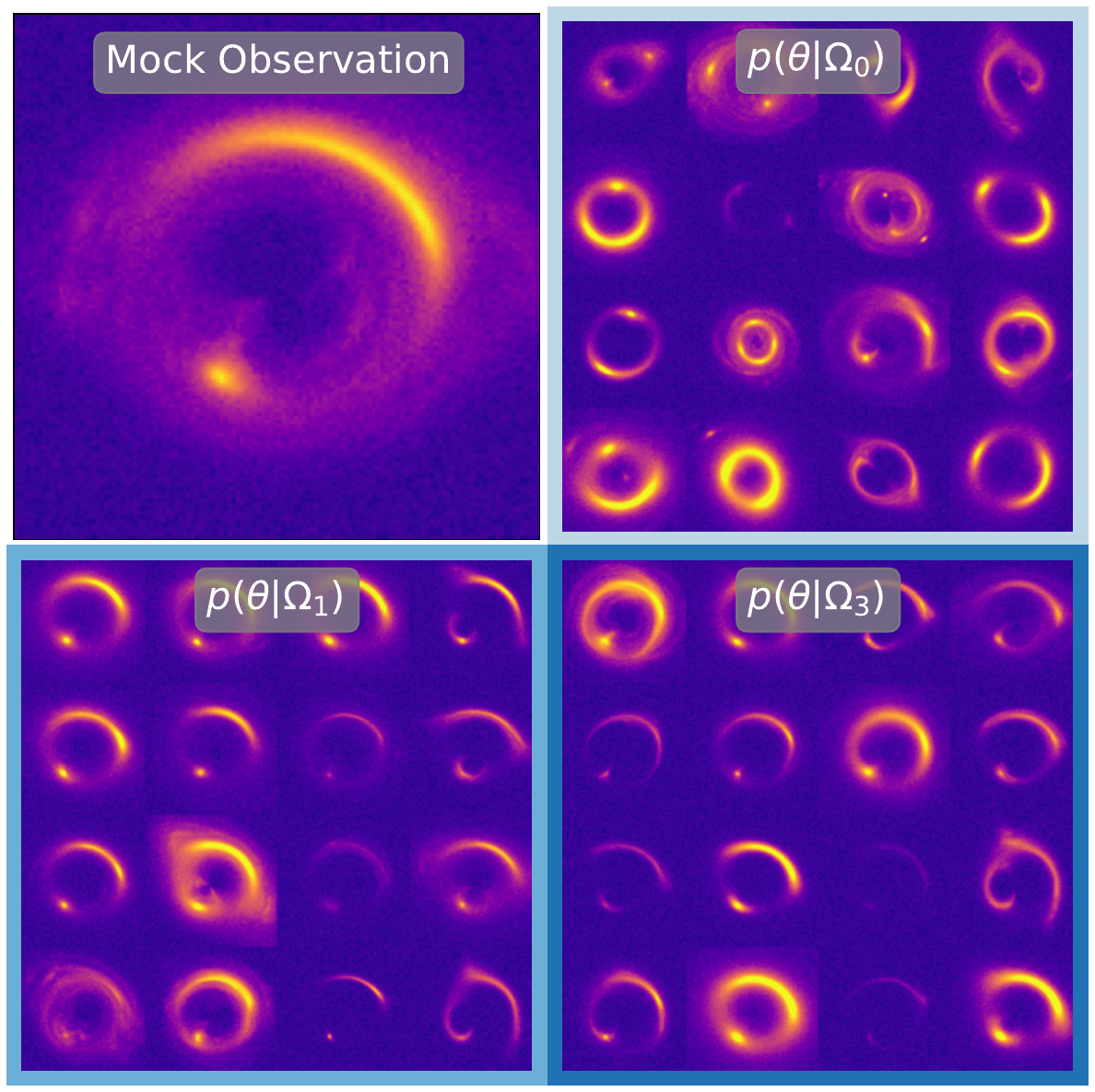}
    \caption{The mock observations (upper left) compared to draws from the prior (upper right, light blue box), the first sequential proposal (lower left, blue box), and the final sequential proposal (lower right, dark blue box).}
    \label{fig:seq_image_comparison}
\end{figure}

To better understand the source of this improvement in performance, we plot the mean loss on the Einstein radius, $\theta_E$, for our 30 mock observations in Figure \ref{fig:sequential_loss_theta_e}. Unlike for $\Sigma_\text{sub}$, the loss for both methods begins decreasing rapidly. Within ten epochs, a model trained on the broad prior can place constraints on the Einstein radius that are meaningfully tighter than the prior. The loss comparison for the remaining lensing parameters, presented in Appendix \ref{app:seq_comp_add_params}, shows a similar trend. In turn, the sequential proposals can substantially narrow the range of lensing parameters being considered. To illustrate this, in Figure \ref{fig:seq_image_comparison}, we show a comparison of one of our mock observations to draw from the broad prior, the first sequential proposal distribution, and the final sequential proposal distribution. Even for the first sequential proposal, the training images produced match the mock observations closely. In short, the sequential approach benefits from constraining the comparatively `easy' lensing parameters early on, thus enabling it to extract the signal on the `difficult' parameter, $\Sigma_\text{sub}$.

\section{Results: Hierarchical Inference Comparison}\label{sec:hier_inf_results}

\begin{figure*}
    \centering
    \begin{subfigure}[t]{0.48\textwidth}
        \centering
        \includegraphics[width=\textwidth]{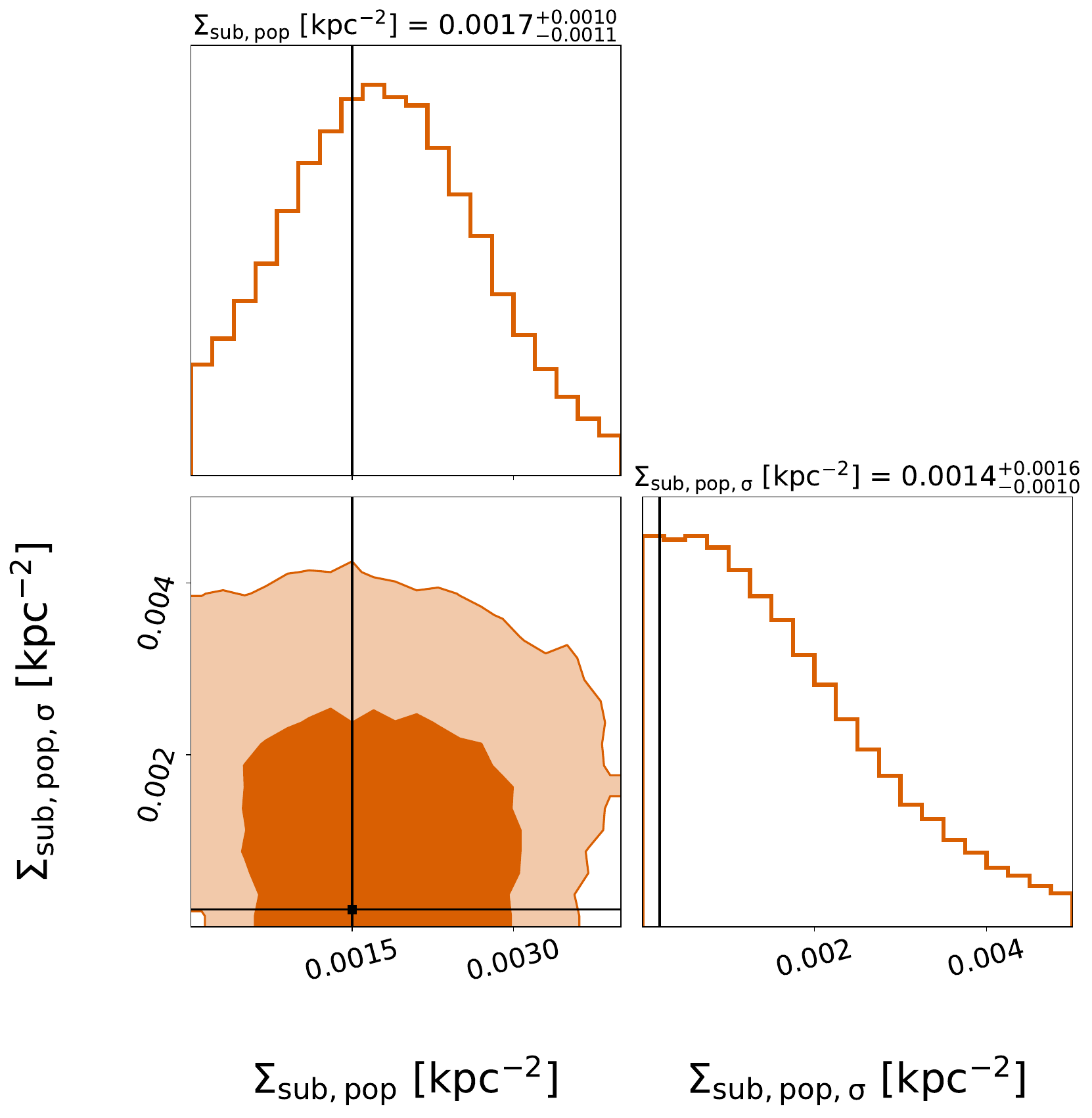}
        \caption{Fiducial model -- 10 lens analysis}
        \label{fig:hier_inf_fid_10}
    \end{subfigure}%
    ~ 
    \begin{subfigure}[t]{0.48\textwidth}
        \centering
        \includegraphics[width=\textwidth]{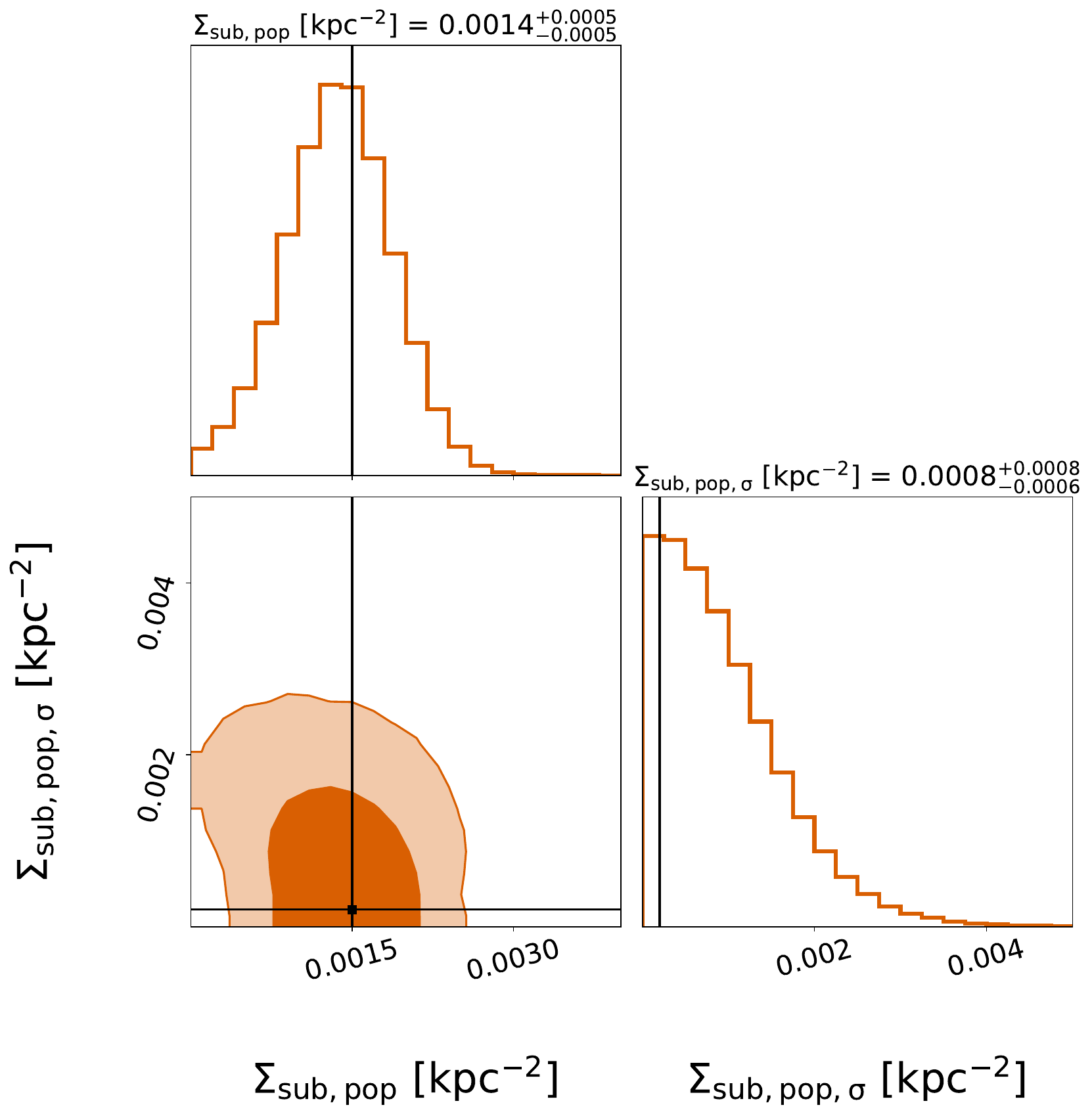}
        \caption{Fiducial model -- 30 lens analysis}
    \end{subfigure} \\
    \vspace{0.5cm}
    \begin{subfigure}[t]{0.48\textwidth}
        \centering
        \includegraphics[width=\textwidth]{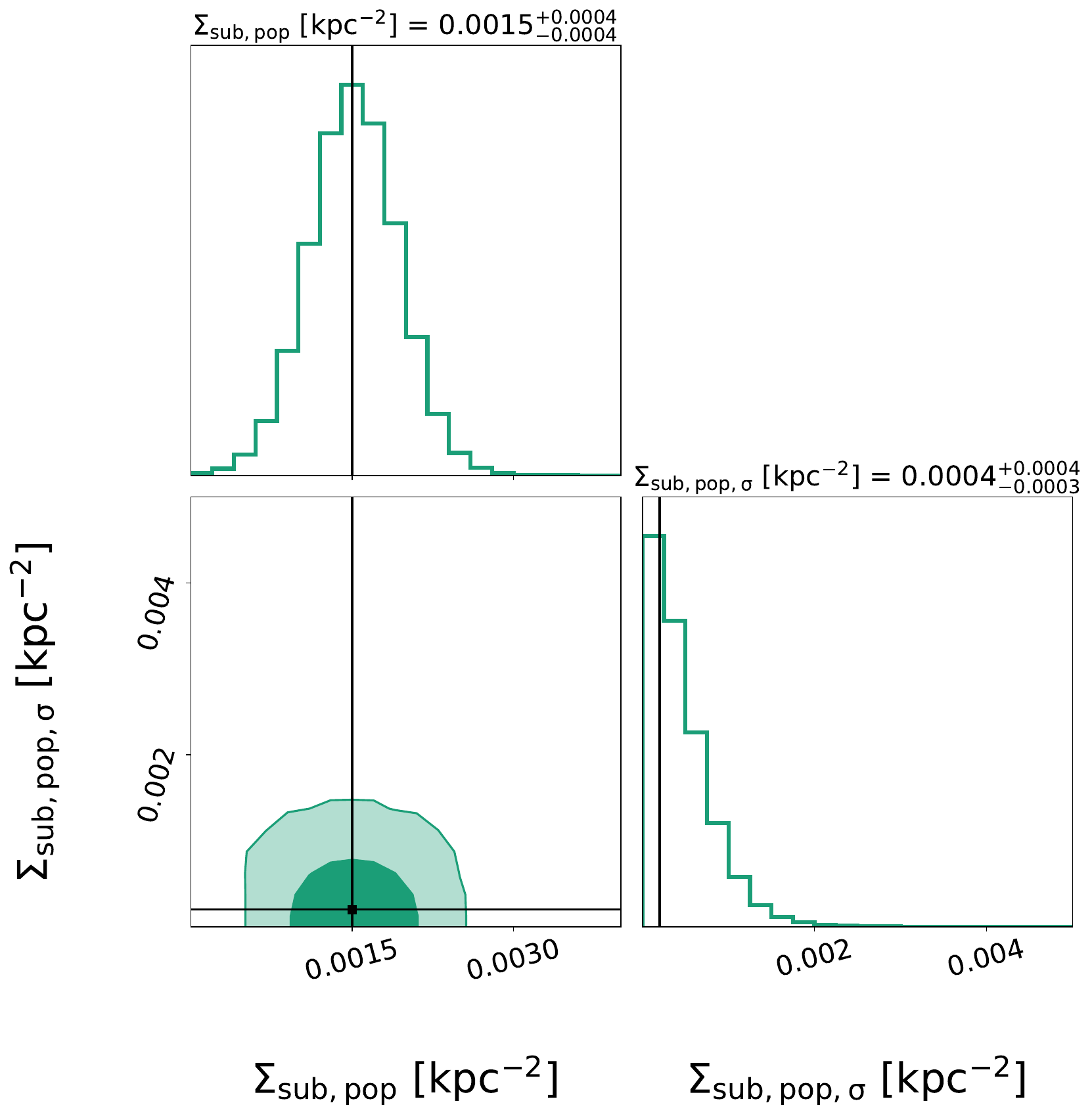}
        \caption{Sequential model -- 10 lens analysis}
    \end{subfigure}%
    ~ 
    \begin{subfigure}[t]{0.48\textwidth}
        \centering
        \includegraphics[width=\textwidth]{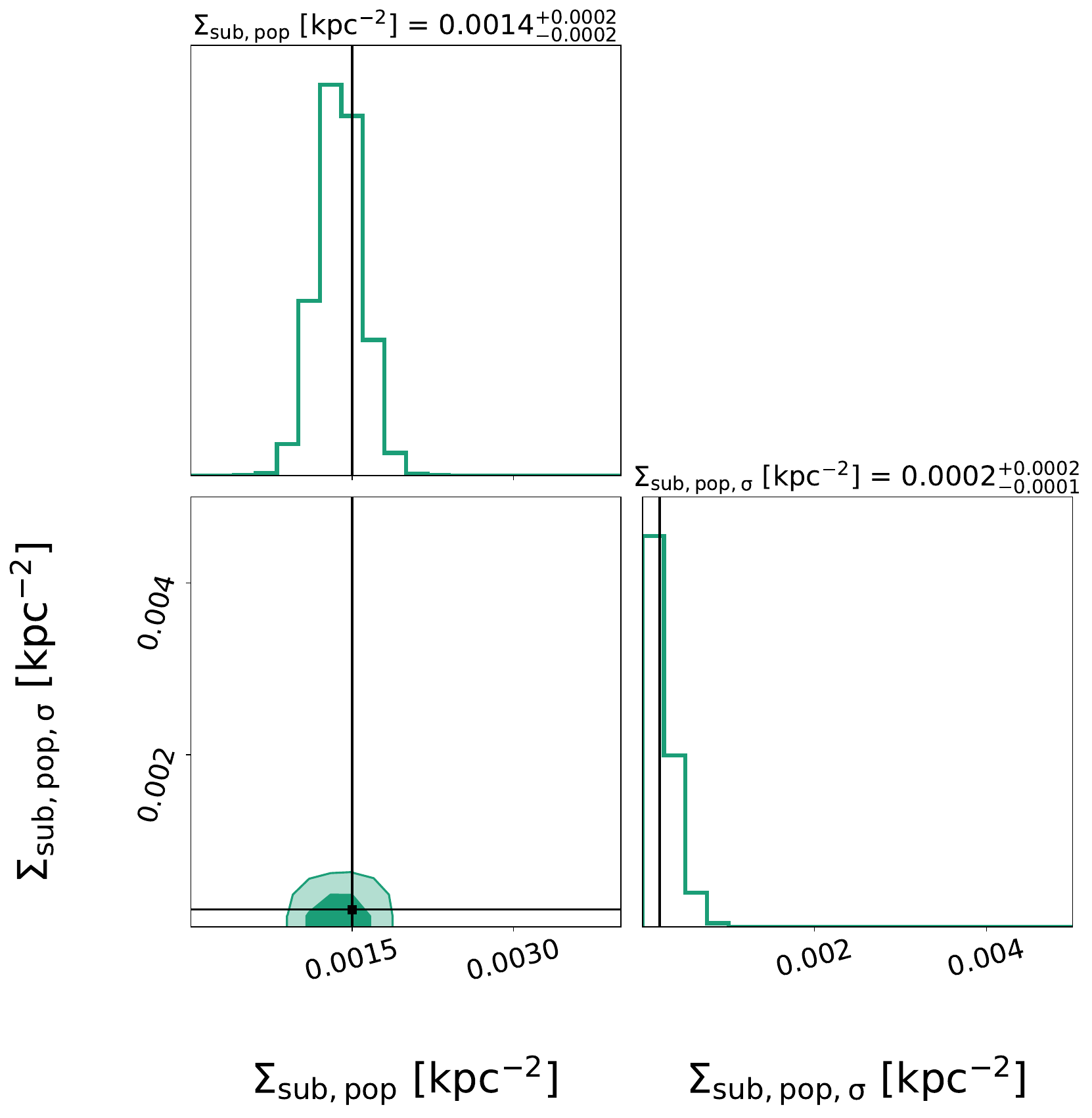}
        \caption{Sequential model -- 30 lens analysis}
    \end{subfigure}
    \caption{Hierarchical inference results on the mean, $\Sigma_\text{sub,pop}$, and scatter, $\Sigma_{\text{sub,pop},\sigma}$, of the population distribution for $\Sigma_\text{sub}$. For the fiducial, NPE approach, (a) shows the constraints with 10 lenses, and (b) shows the constraints with 30 lenses. For the sequential, SNPE approach, (c) shows the constraints with 10 lenses, and (d) shows the constraints with 30 lenses. All of the lenses are drawn from the mock observation set. In each plot, the dark and light contours represent the  68\% and 95\% confidence intervals, respectively. The black dot shows the true value from which the mock observations were generated.}
    \label{fig:hier_inf}
\end{figure*}

Section \ref{sec:seq_inf_results} demonstrates that the sequential SNPE approach achieves a performance on the loss that appears computationally unreachable with the fiducial NPE approach. In this section, we show how those improvements in loss translate to the constraining power of our model. Specifically, we run a hierarchical inference on the mean, $\Sigma_\text{sub,pop}$, and scatter, $\Sigma_{\text{sub,pop},\sigma}$, of the population distribution for $\Sigma_\text{sub}$. For our density estimator, we compare the fiducial model after 500 epochs of training to our 30 sequential models after 26 epochs of training. This allows us to compare the constraining power of the two approaches when using the same total computational resources. For the images, we use the same set of test images that were introduced in Section \ref{sec:seq_inf_results}. We also introduce 90 additional mock observations drawn from the same distribution to better understand the scaling of the fiducial method.

The hierarchical constraints for a set of 10 and 30 lenses are shown in Figure \ref{fig:hier_inf} for the fiducial and sequential approach. All four constraints contain the true population parameters within the 68\% confidence interval. The lower loss achieved by the sequential method translates into much stronger hierarchical constraints. The 10 lens constraint from the sequential models outperforms the 30 lens constraint from the fiducial model, largely due to a significant improvement in the constraining power on $\Sigma_{\text{sub,pop},\sigma}$. None of the four constraints shown demonstrate a bias toward the training prior. In fact, the one-dimensional marginals of the 30 lens constraint using the sequential model disfavor the training prior values of $\Sigma_\text{sub,pop} = 2 \times 10^{-3} \text{kpc}^{-2} $ at three sigma and $\Sigma_{\text{sub,pop},\sigma} = 1.1 \times 10^{-3} \text{kpc}^{-2}$ at five sigma. 

\begin{figure*}
    \centering
	\includegraphics[width=0.8\textwidth]{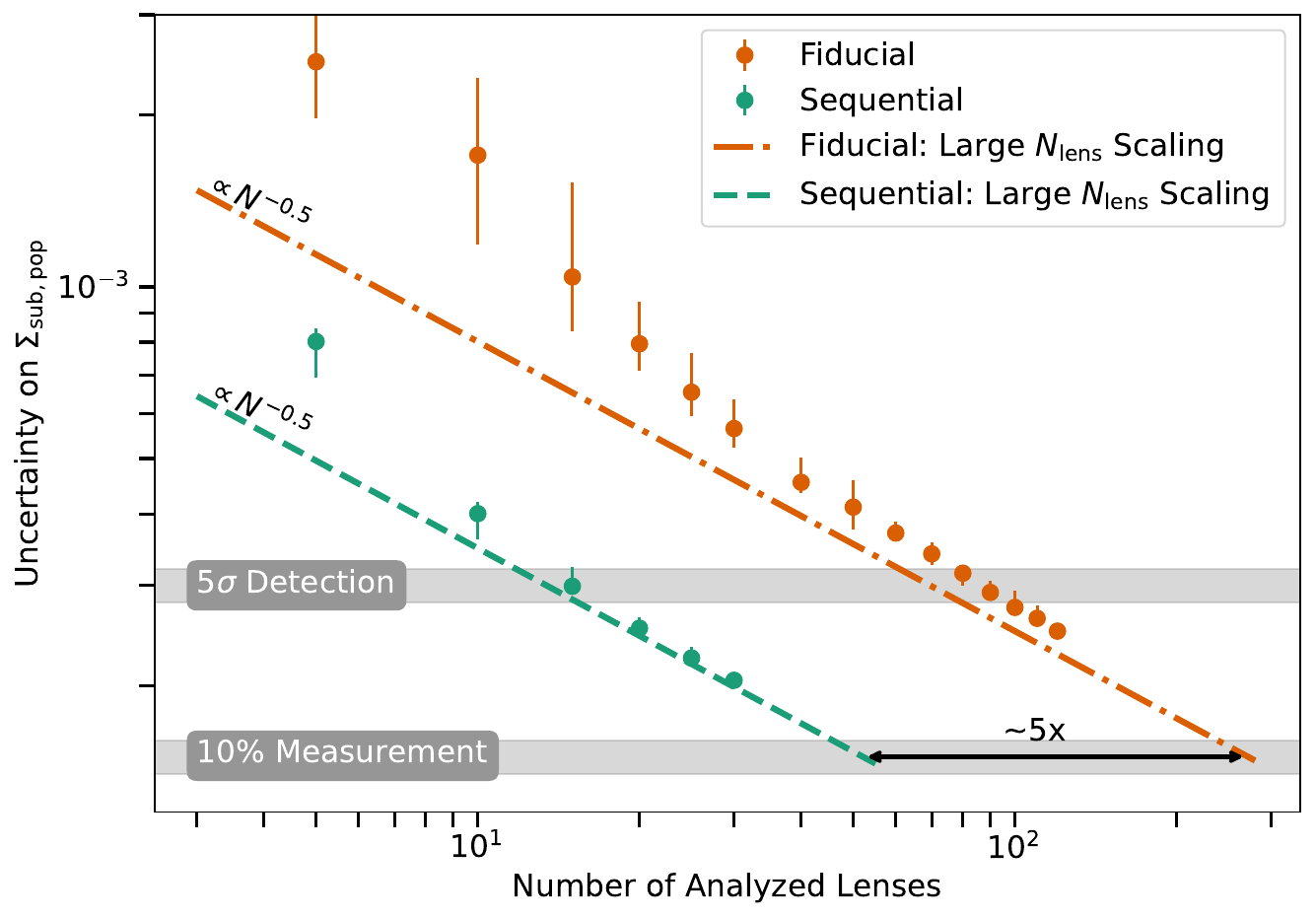}
    \caption{The one-sigma uncertainty on the $\Sigma_\text{sub,pop}$ population constraint as a function of the number of lenses in the analysis for both the fiducial NPE and sequential SNPE approaches. Each point represents the median value from 10 resamplings of the larger mock observation set. The error bars on the points span the 68\% quantiles. The power-law large $N_\text{lens}$ scaling for the fiducial and sequential models are fit to hierarchical analysis where $\Sigma_{\text{sub,pop},\sigma}$ is known (see Appendix \ref{app:large_n_hier_const}). The `$5 \sigma$ Detection' contour corresponds to the one-sigma uncertainty necessary to make a five-sigma detection of a non-zero $\Sigma_{\text{sub,pop}}$. The `$10\%$ Measurement' contour corresponds to a one-sigma uncertainty that is $10\%$ of the true $\Sigma_{\text{sub,pop}}$ value.}
    \label{fig:scaling_hierarchical}
\end{figure*}

In Figure \ref{fig:scaling_hierarchical}, we plot the one-sigma uncertainty on $\Sigma_\text{sub,pop}$ for both the fiducial and sequential methods as a function of the number of lenses in the analysis. Since our sequential analysis can select from 30 lenses and our fiducial analysis can select from 120, we repeat each analysis ten times with a different subsample of lenses. We plot the 68\% quantile and the median produced by the different subsamples. We also introduce a power-law that captures the scaling for large lens samples. The power law is fit to a set of hierarchical analyses where $\Sigma_{\text{sub,pop},\sigma}$ is known (see Appendix \ref{app:large_n_hier_const} for a discussion). On all the samples from five to thirty lenses, the sequential model leads to substantial improvements in constraining power compared to the fiducial model. This statement holds true even when considering the different possible subsamples for each analysis. In other words, the `luckiest' fiducial analysis performs worse than the comparable, `unluckiest' sequential analysis. Even with 120 lenses, the fiducial analysis has weaker constraints than the 30-lens analysis with the sequential models. The scaling on the uncertainty suggests that the fiducial method requires around five times as much data as the sequential analysis to achieve equivalent constraining power on $\Sigma_\text{sub,pop}$.

Figure \ref{fig:scaling_hierarchical} also provides two additional baselines: the sensitivity required to measure $\Sigma_\text{sub,pop}>0$ at five sigma, and an uncertainty equivalent to ten percent of the $\Sigma_\text{sub,pop}$ value. The fiducial model requires around 80 lenses for a five-sigma detection and nearly 300 lenses for a ten percent measurement. By comparison, the sequential models require only 15 lenses for the five sigma measurement and around 55 lenses for the ten percent measurement. By comparison, the Sloan Lens ACS Survey (SLACS) includes roughly 100 lenses measured with the ACS or WFC3 camera on HST \citep{bolton2008slacscomplete,shu2017slacs}. Even if we only use lenses from one of the two cameras, it should be possible to meet both thresholds with the sequential analysis. The fiducial analysis would require nearly three times the SLACS sample to make a ten percent measurement.

\section{Discussion}\label{sec:discussion}

In the previous sections, we compared possible limitations of NPE methods for strong lensing analysis. We then introduced SNPE as a logical extension of the NPE approach and probed its performance on the loss and the final hierarchical constraining power of a set of mock observations. In this section, we will outline some of the limitations of our results and discuss how these limitations can be overcome as we bring our analysis to the data.

One of the primary conclusions of our paper is that the performance of NPE models on strong lensing is primarily limited by the training set. This conclusion motivates the introduction of SNPE, which allows our training images to be drawn from distributions that are denser near the observations. SNPE leads to large performance gains, but it may also alter our initial conclusions. First, with the leap in performance, model size and architecture may become the primary limitations. Both the ResNet 50 architecture and the choice of a diagonal multivariate Gaussian distribution place fundamental limits on the capacity of our model that may no longer be appropriate for our improved precision. Additionally, SNPE introduces several new modeling choices we have not fully optimized for in this work. For example, in our main results we chose to sample a new proposal distribution every ten epochs, and we set that distribution equal to the current posterior. Neither of these choices is a requirement of SNPE (see Appendix \ref{app:seq_proposal}). Similarly, we have not conducted a thorough exploration of the optimal learning rate for sequential inference. As we alter the underlying distribution from which our training samples are drawn, we are also altering the loss landscape our model is optimizing. More nuanced learning rates that account for the transition between proposal distributions could lead to further gains in performance. We leave the exploration of these sequential modeling choices for future work.

Similarly, our current sequential approach may benefit from incorporating additional lensing parameters beyond those modeled in this paper. Two potential areas for improvement are the source light and the observational effects. Currently, our model outputs predictions on the x- and y-coordinate of the source but outputs no information about its luminosity or morphology. Therefore, our sequential proposals draw from the same luminosity and morphology distributions assumed in our broader prior. As we show in Appendix \ref{app:seq_train_overfit}, the most important of these is likely to be the source morphology. Our pipeline does not include morphological parameters that would allow us to modify the morphological distribution of sources in our sequential proposals. However, it should be possible to generate these parameters, either through simple analytic models (i.e.,\ the S\'{e}rsic profile) or through data-driven encodings. Sequentially inferring the point spread function could also benefit our model. For HST measurements, the observational conditions can vary substantially between different pointings. Rather than simply marginalizing over additional uncertainties in the PSF, sky brightness, or zeropoint, we could steer the observational effects of our simulations to be more representative of the specific data. As with the source morphology, an appropriate encoding of the PSF would be required. 

Additionally, while we have made efforts to generate simulations that are representative of the HST data, there are still a few simplifying assumptions we will need to relax for a final analysis. For the PSF, we currently utilize a single, simple Gaussian convolution. In reality, the HST PSFs have longer tails, non-symmetric structures, and vary over time. Sampling from empirical measurements of the HST PSF will help avoid potential bias. Our simulations also do not include lens light or the line-of-sight halos, which can introduce additional uncertainties. The source light morphological distribution also assumes that our lensing sources are represented by the galaxies sampled in COSMOS. Because we cut on well-resolved galaxies, the COSMOS galaxies will be at lower redshifts than the distribution of lens sources. Errors in the underlying morphological distribution of sources may produce systematic biases in the inferred subhalo population. Future work could correct for any misspecification in the population through the hierarchical analysis, but this would require adding a parameterization of the source morphology to our simulation output. Our current simulations also do not include the effects of the HST drizzling pipeline. The drizzling pipeline has three important effects for our analysis: it mitigates the contribution from cosmic rays and hot pixels, it results in an image with a smaller pixel scale, and it produces correlated noise between pixels. Either drizzling has to be included in our simulator, or our analysis has to be run on the raw exposures. We leave these extensions to the simulation tools for future work.

Finally, our current analysis focuses solely on one parameter of the subhalo distribution: the SHMF normalization. Since the focus of our analysis was primarily on methodological comparison, single-parameter inference was an appropriate scope. However, on the data we will want to incorporate additional parameters. These include a low-mass halo cutoff, the power-law slope of the SHMF, and modifications to the SHMF produced by alternatives to CDM. Within CDM, these parameters would enable us to make precise statements about the sensitivity of our model to specific ranges of halo mass. Outside the context of CDM, these parameters would allow us to probe violations of our current dark matter paradigm.

\vspace{0.5em}

\section{Conclusion}\label{sec:conclusion}

We present an application of SNPE to strong lensing that leads to a factor of five improvement over NPE in our ability to constrain the SHMF normalization. By developing a new simulation suite capable of generating strong lensing images in milliseconds, we explore the limitations of the NPE method to our strong lensing problem. In particular, we consider the architecture, optimization schedule, training set size, and computational resources. We find that the size of the training dataset is the principal limitation. Inspired by this result, we introduce SNPE to improve the quality of our training samples. SNPE yields a substantial reduction in the loss on $\Sigma_\text{sub}$ compared to NPE. Our power-law fit to the NPE loss curve suggests that over three orders of magnitude more computational resources would be required to achieve the SNPE results using NPE. The improvements in the loss enabled by SNPE translate well to the population-level constraints on $\Sigma_\text{sub}$: given equivalent computational resources, the NPE approach requires five times as much data to reach the same constraining power as the SNPE approach. From our mock observations, we conclude that the current SLACS sample of HST strong lenses is sufficient to yield both a five sigma detection and a ten percent measurement of a mean SHMF normalization of $\Sigma_\text{sub,pop} = 1.5 \times 10^{-3} \text{kpc}^{-2}$.

Previous work has shown that simulation-based inference can robustly infer the subhalo populations underlying strong gravitational lenses. Halos in this low-mass regime are a powerful probe of dark matter physics but have been historically difficult to measure through luminous tracers. While the NPE results are promising, they have also demonstrated limited constraining power on individual lenses; substantive constraints only emerge when combining dozens or hundreds of lenses. In this work, we demonstrate that these weak constraints are driven entirely by limitations in our methodology rather than any fundamental property of the data. Given our control over the proposal distribution in SBI, we show that sequential methods can effectively overcome limitations in the training set size without incurring untenable computational costs. While this work focuses on strong lensing, we believe that our results make a poignant argument for integrating training set generation and parameter inference in any cosmological analysis leveraging SBI.

\section*{Acknowledgments}

We would like to thank Phil Marshall for numerous helpful discussions during the development of this work, and Ethan Nadler for helpful comments on a draft. SWC was supported by NSF Award DGE-1656518 and a Stanford Data Science Fellowship while conducting this work. SWC also developed the initial \textsc{paltax} codebase as part of a research internship at Google Research Brain Team\footnote{Now a part of Google DeepMind.} which was publicly released following the internship\footnote{\url{https://github.com/google-research/google-research/tree/master/jaxstronomy}}. SB is partially supported by NASA Roman WFS Award No: 80NSSC24K0095.

This work received support from the Kavli Institute for Particle Astrophysics and Cosmology at Stanford and SLAC through a Kavli Fellowship to JA and from the U.S. Department of Energy under contract number DE-AC02-76SF00515 to SLAC National Accelerator Laboratory.

\bibliography{main}

\begin{thebibliography}{}
\expandafter\ifx\csname natexlab\endcsname\relax\def\natexlab#1{#1}\fi
\providecommand{\url}[1]{\href{#1}{#1}}
\providecommand{\dodoi}[1]{doi:~\href{http://doi.org/#1}{\nolinkurl{#1}}}
\providecommand{\doeprint}[1]{\href{http://ascl.net/#1}{\nolinkurl{http://ascl.net/#1}}}
\providecommand{\doarXiv}[1]{\href{https://arxiv.org/abs/#1}{\nolinkurl{https://arxiv.org/abs/#1}}}

\bibitem[{{Aganze} {et~al.}(2024){Aganze}, {Pearson}, {Starkenburg},
  {Contardo}, {Johnston}, {Tavangar}, {Price-Whelan}, \&
  {Burgasser}}]{roman2024streamforecast}
{Aganze}, C., {Pearson}, S., {Starkenburg}, T., {et~al.} 2024, \apj, 962, 151,
  \dodoi{10.3847/1538-4357/ad159c}

\bibitem[{{Anau Montel} {et~al.}(2023){Anau Montel}, {Coogan}, {Correa},
  {Karchev}, \& {Weniger}}]{noemi2023wdm}
{Anau Montel}, N., {Coogan}, A., {Correa}, C., {Karchev}, K., \& {Weniger}, C.
  2023, \mnras, 518, 2746, \dodoi{10.1093/mnras/stac3215}

\bibitem[{{Banik} {et~al.}(2021){Banik}, {Bovy}, {Bertone}, {Erkal}, \& {de
  Boer}}]{banik2021}
{Banik}, N., {Bovy}, J., {Bertone}, G., {Erkal}, D., \& {de Boer}, T.~J.~L.
  2021, \mnras, 502, 2364, \dodoi{10.1093/mnras/stab210}

\bibitem[{{Barkana}(1998)}]{barkana1998fast}
{Barkana}, R. 1998, \apj, 502, 531, \dodoi{10.1086/305950}

\bibitem[{{Bechtol} {et~al.}(2022){Bechtol}, {Birrer}, {Cyr-Racine}, {Schutz},
  {Adhikari}, {Amin}, {Banerjee}, {Bird}, {Blinov}, {Boddy}, {Boehm}, {Bundy},
  {Buschmann}, {Chakrabarti}, {Curtin}, {Dai}, {Drlica-Wagner}, {Dvorkin},
  {Erickcek}, {Gilman}, {Heeba}, {Kim}, {Ir{\v{s}}i{\v{c}}}, {Leauthaud},
  {Lovell}, {Luki{\'c}}, {Mao}, {Mau}, {Mitridate}, {Mocz}, {Mu{\~n}oz},
  {Nadler}, {Peter}, {Price-Whelan}, {Robertson}, {Sabti}, {Sehgal}, {Shipp},
  {Simon}, {Singh}, {Van Tilburg}, {Wechsler}, {Widmark}, \&
  {Yu}}]{snowmass2021dm_from_halo}
{Bechtol}, K., {Birrer}, S., {Cyr-Racine}, F.-Y., {et~al.} 2022, arXiv
  e-prints, arXiv:2203.07354, \dodoi{10.48550/arXiv.2203.07354}

\bibitem[{{Birrer} \& {Amara}(2018)}]{birrer2018lenstronomy}
{Birrer}, S., \& {Amara}, A. 2018, Physics of the Dark Universe, 22, 189,
  \dodoi{10.1016/j.dark.2018.11.002}

\bibitem[{{Birrer} {et~al.}(2017{\natexlab{a}}){Birrer}, {Amara}, \&
  {Refregier}}]{birrer2017lensing}
{Birrer}, S., {Amara}, A., \& {Refregier}, A. 2017{\natexlab{a}}, \jcap, 2017,
  037, \dodoi{10.1088/1475-7516/2017/05/037}

\bibitem[{{Birrer} {et~al.}(2017{\natexlab{b}}){Birrer}, {Welschen}, {Amara},
  \& {Refregier}}]{birrer2017line}
{Birrer}, S., {Welschen}, C., {Amara}, A., \& {Refregier}, A.
  2017{\natexlab{b}}, \jcap, 2017, 049, \dodoi{10.1088/1475-7516/2017/04/049}

\bibitem[{{Birrer} {et~al.}(2021){Birrer}, {Shajib}, {Gilman}, {Galan},
  {Aalbers}, {Millon}, {Morgan}, {Pagano}, {Park}, {Teodori}, {Tessore},
  {Ueland}, {Van de Vyvere}, {Wagner-Carena}, {Wempe}, {Yang}, {Ding},
  {Schmidt}, {Sluse}, {Zhang}, \& {Amara}}]{birrer2021lenstronomy}
{Birrer}, S., {Shajib}, A., {Gilman}, D., {et~al.} 2021, The Journal of Open
  Source Software, 6, 3283, \dodoi{10.21105/joss.03283}

\bibitem[{{Bode} {et~al.}(2001){Bode}, {Ostriker}, \& {Turok}}]{bode2001halo}
{Bode}, P., {Ostriker}, J.~P., \& {Turok}, N. 2001, \apj, 556, 93,
  \dodoi{10.1086/321541}

\bibitem[{{Bolton} {et~al.}(2008){Bolton}, {Burles}, {Koopmans}, {Treu},
  {Gavazzi}, {Moustakas}, {Wayth}, \& {Schlegel}}]{bolton2008slacscomplete}
{Bolton}, A.~S., {Burles}, S., {Koopmans}, L. V.~E., {et~al.} 2008, \apj, 682,
  964, \dodoi{10.1086/589327}

\bibitem[{Bradbury {et~al.}(2018)Bradbury, Frostig, Hawkins, Johnson, Leary,
  Maclaurin, Necula, Paszke, Vander{P}las, Wanderman-{M}ilne, \&
  Zhang}]{jax2018github}
Bradbury, J., Frostig, R., Hawkins, P., {et~al.} 2018.
\newblock \url{http://github.com/google/jax}

\bibitem[{{Brehmer} {et~al.}(2019){Brehmer}, {Mishra-Sharma}, {Hermans},
  {Louppe}, \& {Cranmer}}]{brehmer2019mining}
{Brehmer}, J., {Mishra-Sharma}, S., {Hermans}, J., {Louppe}, G., \& {Cranmer},
  K. 2019, \apj, 886, 49, \dodoi{10.3847/1538-4357/ab4c41}

\bibitem[{{Buckley} \& {Peter}(2018)}]{buckley2018gravitational}
{Buckley}, M.~R., \& {Peter}, A. H.~G. 2018, \physrep, 761, 1,
  \dodoi{10.1016/j.physrep.2018.07.003}

\bibitem[{{Bullock} \& {Boylan-Kolchin}(2017)}]{bullock2017small}
{Bullock}, J.~S., \& {Boylan-Kolchin}, M. 2017, \araa, 55, 343,
  \dodoi{10.1146/annurev-astro-091916-055313}

\bibitem[{{Calamida} {et~al.}(2021){Calamida}, {Mack}, {Medina}, {Shanahan},
  {Bajaj}, \& {Deustua}}]{calamida2021new}
{Calamida}, A., {Mack}, J., {Medina}, J., {et~al.} 2021, {New time-dependent
  WFC3 UVIS inverse sensitivities}, Instrument Science Report WFC3 2021-4, 33
  pages

\bibitem[{{Carlson} {et~al.}(1992){Carlson}, {Machacek}, \&
  {Hall}}]{Carlson1992}
{Carlson}, E.~D., {Machacek}, M.~E., \& {Hall}, L.~J. 1992, \apj, 398, 43,
  \dodoi{10.1086/171833}

\bibitem[{{Cranmer} {et~al.}(2020){Cranmer}, {Brehmer}, \& {Louppe}}]{sbi}
{Cranmer}, K., {Brehmer}, J., \& {Louppe}, G. 2020, Proceedings of the National
  Academy of Science, 117, 30055, \dodoi{10.1073/pnas.1912789117}

\bibitem[{{Cyr-Racine} {et~al.}(2016){Cyr-Racine}, {Moustakas}, {Keeton},
  {Sigurdson}, \& {Gilman}}]{cyr2016dark}
{Cyr-Racine}, F.-Y., {Moustakas}, L.~A., {Keeton}, C.~R., {Sigurdson}, K., \&
  {Gilman}, D.~A. 2016, \prd, 94, 043505, \dodoi{10.1103/PhysRevD.94.043505}

\bibitem[{{Dalal} \& {Kochanek}(2002)}]{dalal2002direct}
{Dalal}, N., \& {Kochanek}, C.~S. 2002, \apj, 572, 25, \dodoi{10.1086/340303}

\bibitem[{{Diaz Rivero} {et~al.}(2018){Diaz Rivero}, {Cyr-Racine}, \&
  {Dvorkin}}]{rivero2018power}
{Diaz Rivero}, A., {Cyr-Racine}, F.-Y., \& {Dvorkin}, C. 2018, \prd, 97,
  023001, \dodoi{10.1103/PhysRevD.97.023001}

\bibitem[{{D{\'\i}az Rivero} {et~al.}(2018){D{\'\i}az Rivero}, {Dvorkin},
  {Cyr-Racine}, {Zavala}, \& {Vogelsberger}}]{rivero2018gravitational}
{D{\'\i}az Rivero}, A., {Dvorkin}, C., {Cyr-Racine}, F.-Y., {Zavala}, J., \&
  {Vogelsberger}, M. 2018, \prd, 98, 103517, \dodoi{10.1103/PhysRevD.98.103517}

\bibitem[{{Diemer}(2018)}]{diemer2018colossus}
{Diemer}, B. 2018, \apjs, 239, 35, \dodoi{10.3847/1538-4365/aaee8c}

\bibitem[{{Foreman-Mackey} {et~al.}(2014){Foreman-Mackey}, {Hogg}, \&
  {Morton}}]{foreman-mackey2014}
{Foreman-Mackey}, D., {Hogg}, D.~W., \& {Morton}, T.~D. 2014, \apj, 795, 64,
  \dodoi{10.1088/0004-637X/795/1/64}

\bibitem[{{Gennaro}(2018)}]{wfc3_datahandbook}
{Gennaro}, M. 2018, in WFC3 Data Handbook v. 4, Vol.~4, 4

\bibitem[{{Gilman} {et~al.}(2020{\natexlab{a}}){Gilman}, {Birrer},
  {Nierenberg}, {Treu}, {Du}, \& {Benson}}]{gilman2020warm}
{Gilman}, D., {Birrer}, S., {Nierenberg}, A., {et~al.} 2020{\natexlab{a}},
  \mnras, 491, 6077, \dodoi{10.1093/mnras/stz3480}

\bibitem[{{Gilman} {et~al.}(2020{\natexlab{b}}){Gilman}, {Du}, {Benson},
  {Birrer}, {Nierenberg}, \& {Treu}}]{gilman2020constraints}
{Gilman}, D., {Du}, X., {Benson}, A., {et~al.} 2020{\natexlab{b}}, \mnras, 492,
  L12, \dodoi{10.1093/mnrasl/slz173}

\bibitem[{{Goodman} \& {Weare}(2010)}]{goodman2010ensemble}
{Goodman}, J., \& {Weare}, J. 2010, Communications in Applied Mathematics and
  Computational Science, 5, 65, \dodoi{10.2140/camcos.2010.5.65}

\bibitem[{{Green} {et~al.}(2004){Green}, {Hofmann}, \&
  {Schwarz}}]{green2004power}
{Green}, A.~M., {Hofmann}, S., \& {Schwarz}, D.~J. 2004, \mnras, 353, L23,
  \dodoi{10.1111/j.1365-2966.2004.08232.x}

\bibitem[{Greenberg {et~al.}(2019)Greenberg, Nonnenmacher, \&
  Macke}]{greenberg2019automatic}
Greenberg, D., Nonnenmacher, M., \& Macke, J. 2019, in International Conference
  on Machine Learning, PMLR, 2404--2414.
\newblock \url{http://proceedings.mlr.press/v97/greenberg19a/greenberg19a.pdf}

\bibitem[{He {et~al.}(2016)He, Zhang, Ren, \& Sun}]{he2016deep}
He, K., Zhang, X., Ren, S., \& Sun, J. 2016, in Proceedings of the IEEE
  conference on computer vision and pattern recognition, 770--778.
\newblock
  \url{https://www.cv-foundation.org/openaccess/content_cvpr_2016/papers/He_Deep_Residual_Learning_CVPR_2016_paper.pdf}

\bibitem[{{He} {et~al.}(2018){He}, {Zhang}, {Zhang}, {Zhang}, {Xie}, \&
  {Li}}]{he2019bag}
{He}, T., {Zhang}, Z., {Zhang}, H., {et~al.} 2018, arXiv e-prints,
  arXiv:1812.01187, \dodoi{10.48550/arXiv.1812.01187}

\bibitem[{{Hezaveh} {et~al.}(2016{\natexlab{a}}){Hezaveh}, {Dalal}, {Holder},
  {Kisner}, {Kuhlen}, \& {Perreault Levasseur}}]{hezaveh2016measuring}
{Hezaveh}, Y., {Dalal}, N., {Holder}, G., {et~al.} 2016{\natexlab{a}}, \jcap,
  2016, 048, \dodoi{10.1088/1475-7516/2016/11/048}

\bibitem[{{Hezaveh} {et~al.}(2013){Hezaveh}, {Dalal}, {Holder}, {Kuhlen},
  {Marrone}, {Murray}, \& {Vieira}}]{hezaveh2013dark}
---. 2013, \apj, 767, 9, \dodoi{10.1088/0004-637X/767/1/9}

\bibitem[{{Hezaveh} {et~al.}(2016{\natexlab{b}}){Hezaveh}, {Dalal}, {Marrone},
  {Mao}, {Morningstar}, {Wen}, {Blandford}, {Carlstrom}, {Fassnacht}, {Holder},
  {Kemball}, {Marshall}, {Murray}, {Perreault Levasseur}, {Vieira}, \&
  {Wechsler}}]{hezaveh2016detection}
{Hezaveh}, Y.~D., {Dalal}, N., {Marrone}, D.~P., {et~al.} 2016{\natexlab{b}},
  \apj, 823, 37, \dodoi{10.3847/0004-637X/823/1/37}

\bibitem[{{Hu} {et~al.}(2000){Hu}, {Barkana}, \& {Gruzinov}}]{hufidm2000}
{Hu}, W., {Barkana}, R., \& {Gruzinov}, A. 2000, \prl, 85, 1158,
  \dodoi{10.1103/PhysRevLett.85.1158}

\bibitem[{{Hui} {et~al.}(2017){Hui}, {Ostriker}, {Tremaine}, \&
  {Witten}}]{hui2017}
{Hui}, L., {Ostriker}, J.~P., {Tremaine}, S., \& {Witten}, E. 2017, \prd, 95,
  043541, \dodoi{10.1103/PhysRevD.95.043541}

\bibitem[{{Kaplinghat}(2005)}]{kaplinghat2005dark}
{Kaplinghat}, M. 2005, \prd, 72, 063510, \dodoi{10.1103/PhysRevD.72.063510}

\bibitem[{{Keeton} {et~al.}(1997){Keeton}, {Kochanek}, \&
  {Seljak}}]{keeton1997shear}
{Keeton}, C.~R., {Kochanek}, C.~S., \& {Seljak}, U. 1997, \apj, 482, 604,
  \dodoi{10.1086/304172}

\bibitem[{Kingma \& Ba(2015)}]{kingma2015adam}
Kingma, D.~P., \& Ba, J. 2015, in 3rd International Conference on Learning
  Representations, {ICLR} 2015, San Diego, CA, USA, May 7-9, 2015, Conference
  Track Proceedings, ed. Y.~Bengio \& Y.~LeCun.
\newblock \url{http://arxiv.org/abs/1412.6980}

\bibitem[{{Koekemoer} {et~al.}(2007){Koekemoer}, {Aussel}, {Calzetti}, {Capak},
  {Giavalisco}, {Kneib}, {Leauthaud}, {Le F{\`e}vre}, {McCracken}, {Massey},
  {Mobasher}, {Rhodes}, {Scoville}, \& {Shopbell}}]{koekemoer2007cosmos}
{Koekemoer}, A.~M., {Aussel}, H., {Calzetti}, D., {et~al.} 2007, \apjs, 172,
  196, \dodoi{10.1086/520086}

\bibitem[{{Koopmans}(2005)}]{koopmans2005gravitational}
{Koopmans}, L.~V.~E. 2005, \mnras, 363, 1136,
  \dodoi{10.1111/j.1365-2966.2005.09523.x}

\bibitem[{{Kormann} {et~al.}(1994){Kormann}, {Schneider}, \&
  {Bartelmann}}]{kormann1994isothermal}
{Kormann}, R., {Schneider}, P., \& {Bartelmann}, M. 1994, \aap, 284, 285

\bibitem[{{Kummer} {et~al.}(2018){Kummer}, {Kahlhoefer}, \&
  {Schmidt-Hoberg}}]{kummer2018}
{Kummer}, J., {Kahlhoefer}, F., \& {Schmidt-Hoberg}, K. 2018, \mnras, 474, 388,
  \dodoi{10.1093/mnras/stx2715}

\bibitem[{Loshchilov \& Hutter(2017)}]{loshchilov2017sgdr}
Loshchilov, I., \& Hutter, F. 2017, in International Conference on Learning
  Representations.
\newblock \url{https://openreview.net/forum?id=Skq89Scxx}

\bibitem[{{Lovell} {et~al.}(2014){Lovell}, {Frenk}, {Eke}, {Jenkins}, {Gao}, \&
  {Theuns}}]{lovell2014}
{Lovell}, M.~R., {Frenk}, C.~S., {Eke}, V.~R., {et~al.} 2014, \mnras, 439, 300,
  \dodoi{10.1093/mnras/stt2431}

\bibitem[{{Lueckmann} {et~al.}(2017){Lueckmann}, {Goncalves}, {Bassetto},
  {{\"O}cal}, {Nonnenmacher}, \& {Macke}}]{lueckmann2017flexible}
{Lueckmann}, J.-M., {Goncalves}, P.~J., {Bassetto}, G., {et~al.} 2017, arXiv
  e-prints, arXiv:1711.01861, \dodoi{10.48550/arXiv.1711.01861}

\bibitem[{{Mandelbaum} {et~al.}(2012){Mandelbaum}, {Hirata}, {Leauthaud},
  {Massey}, \& {Rhodes}}]{mandelbaum2012precision}
{Mandelbaum}, R., {Hirata}, C.~M., {Leauthaud}, A., {Massey}, R.~J., \&
  {Rhodes}, J. 2012, \mnras, 420, 1518,
  \dodoi{10.1111/j.1365-2966.2011.20138.x}

\bibitem[{{Mandelbaum} {et~al.}(2014){Mandelbaum}, {Rowe}, {Bosch}, {Chang},
  {Courbin}, {Gill}, {Jarvis}, {Kannawadi}, {Kacprzak}, {Lackner}, {Leauthaud},
  {Miyatake}, {Nakajima}, {Rhodes}, {Simet}, {Zuntz}, {Armstrong}, {Bridle},
  {Coupon}, {Dietrich}, {Gentile}, {Heymans}, {Jurling}, {Kent}, {Kirkby},
  {Margala}, {Massey}, {Melchior}, {Peterson}, {Roodman}, \&
  {Schrabback}}]{mandelbaum2014third}
{Mandelbaum}, R., {Rowe}, B., {Bosch}, J., {et~al.} 2014, \apjs, 212, 5,
  \dodoi{10.1088/0067-0049/212/1/5}

\bibitem[{{Mao} \& {Schneider}(1998)}]{mao1998evidence}
{Mao}, S., \& {Schneider}, P. 1998, \mnras, 295, 587,
  \dodoi{10.1046/j.1365-8711.1998.01319.x}

\bibitem[{{Mohamed} \& {Lakshminarayanan}(2016)}]{mohamed2016learning}
{Mohamed}, S., \& {Lakshminarayanan}, B. 2016, arXiv e-prints,
  arXiv:1610.03483, \dodoi{10.48550/arXiv.1610.03483}

\bibitem[{{Moore} {et~al.}(1999){Moore}, {Ghigna}, {Governato}, {Lake},
  {Quinn}, {Stadel}, \& {Tozzi}}]{moore1999dark}
{Moore}, B., {Ghigna}, S., {Governato}, F., {et~al.} 1999, \apjl, 524, L19,
  \dodoi{10.1086/312287}

\bibitem[{{Moustakas} \& {Metcalf}(2003)}]{moustakas2003detecting}
{Moustakas}, L.~A., \& {Metcalf}, R.~B. 2003, \mnras, 339, 607,
  \dodoi{10.1046/j.1365-8711.2003.06055.x}

\bibitem[{{Nadler} {et~al.}(2021){Nadler}, {Birrer}, {Gilman}, {Wechsler},
  {Du}, {Benson}, {Nierenberg}, \& {Treu}}]{nadler2021dark}
{Nadler}, E.~O., {Birrer}, S., {Gilman}, D., {et~al.} 2021, \apj, 917, 7,
  \dodoi{10.3847/1538-4357/abf9a3}

\bibitem[{{Nadler} {et~al.}(2023){Nadler}, {Mansfield}, {Wang}, {Du},
  {Adhikari}, {Banerjee}, {Benson}, {Darragh-Ford}, {Mao}, {Wagner-Carena},
  {Wechsler}, \& {Wu}}]{nadler2013zoom}
{Nadler}, E.~O., {Mansfield}, P., {Wang}, Y., {et~al.} 2023, \apj, 945, 159,
  \dodoi{10.3847/1538-4357/acb68c}

\bibitem[{{Navarro} {et~al.}(1996){Navarro}, {Frenk}, \&
  {White}}]{navarro1996structure}
{Navarro}, J.~F., {Frenk}, C.~S., \& {White}, S. D.~M. 1996, \apj, 462, 563,
  \dodoi{10.1086/177173}

\bibitem[{{Navarro} {et~al.}(1997){Navarro}, {Frenk}, \&
  {White}}]{navarro1997universal}
---. 1997, \apj, 490, 493, \dodoi{10.1086/304888}

\bibitem[{{O'Riordan} {et~al.}(2023){O'Riordan}, {Despali}, {Vegetti},
  {Lovell}, \& {Molin{\'e}}}]{oriordan2023sensitivity}
{O'Riordan}, C.~M., {Despali}, G., {Vegetti}, S., {Lovell}, M.~R., \&
  {Molin{\'e}}, {\'A}. 2023, \mnras, 521, 2342, \dodoi{10.1093/mnras/stad650}

\bibitem[{Papamakarios \& Murray(2016)}]{papamakarios2016fast}
Papamakarios, G., \& Murray, I. 2016, Advances in neural information processing
  systems, 29.
\newblock
  \url{https://proceedings.neurips.cc/paper_files/paper/2016/file/6aca97005c68f1206823815f66102863-Paper.pdf}

\bibitem[{Papamakarios {et~al.}(2019)Papamakarios, Sterratt, \&
  Murray}]{papamakarios2019sequential}
Papamakarios, G., Sterratt, D., \& Murray, I. 2019, in The 22nd International
  Conference on Artificial Intelligence and Statistics, PMLR, 837--848.
\newblock
  \url{http://proceedings.mlr.press/v89/papamakarios19a/papamakarios19a.pdf}

\bibitem[{{Park} {et~al.}(2023){Park}, {Birrer}, {Ueland}, {Cranmer},
  {Agnello}, {Wagner-Carena}, {Marshall}, {Roodman}, \& {the LSST Dark Energy
  Science Collaboration}}]{park2023kappa}
{Park}, J.~W., {Birrer}, S., {Ueland}, M., {et~al.} 2023, \apj, 953, 178,
  \dodoi{10.3847/1538-4357/acdc25}

\bibitem[{{Pearson} {et~al.}(2021){Pearson}, {Maresca}, {Li}, \&
  {Dye}}]{pearson2021strong}
{Pearson}, J., {Maresca}, J., {Li}, N., \& {Dye}, S. 2021, \mnras, 505, 4362,
  \dodoi{10.1093/mnras/stab1547}

\bibitem[{{Perreault Levasseur} {et~al.}(2017){Perreault Levasseur}, {Hezaveh},
  \& {Wechsler}}]{perreault2017uncertainties}
{Perreault Levasseur}, L., {Hezaveh}, Y.~D., \& {Wechsler}, R.~H. 2017, \apjl,
  850, L7, \dodoi{10.3847/2041-8213/aa9704}

\bibitem[{{Rogers} \& {Peiris}(2021)}]{rogers2021strong}
{Rogers}, K.~K., \& {Peiris}, H.~V. 2021, \prl, 126, 071302,
  \dodoi{10.1103/PhysRevLett.126.071302}

\bibitem[{Rubin(1984)}]{rubin1984bayesianly}
Rubin, D.~B. 1984, The Annals of Statistics, 1151

\bibitem[{{Rudakovskyi} {et~al.}(2021){Rudakovskyi}, {Mesinger}, {Savchenko},
  \& {Gillet}}]{rudakovskyi2021constraints}
{Rudakovskyi}, A., {Mesinger}, A., {Savchenko}, D., \& {Gillet}, N. 2021,
  \mnras, 507, 3046, \dodoi{10.1093/mnras/stab2333}

\bibitem[{{Ryon}(2021)}]{acs_inshandbook}
{Ryon}, J.~E. 2021, in ACS Instrument Handbook for Cycle 29 v. 20.0, Vol.~20,
  20

\bibitem[{{Schive} {et~al.}(2016){Schive}, {Chiueh}, {Broadhurst}, \&
  {Huang}}]{schive2016}
{Schive}, H.-Y., {Chiueh}, T., {Broadhurst}, T., \& {Huang}, K.-W. 2016, \apj,
  818, 89, \dodoi{10.3847/0004-637X/818/1/89}

\bibitem[{{Schneider} {et~al.}(2012){Schneider}, {Smith}, {Macci{\`o}}, \&
  {Moore}}]{Schneider2012}
{Schneider}, A., {Smith}, R.~E., {Macci{\`o}}, A.~V., \& {Moore}, B. 2012,
  \mnras, 424, 684, \dodoi{10.1111/j.1365-2966.2012.21252.x}

\bibitem[{{Seng{\"u}l} {et~al.}(2022){Seng{\"u}l}, {Dvorkin}, {Ostdiek}, \&
  {Tsang}}]{ccaugan2021substructure}
{Seng{\"u}l}, A.~{\c{C}}., {Dvorkin}, C., {Ostdiek}, B., \& {Tsang}, A. 2022,
  \mnras, 515, 4391, \dodoi{10.1093/mnras/stac1967}

\bibitem[{{Shi} \& {Fuller}(1999)}]{shi1999}
{Shi}, X., \& {Fuller}, G.~M. 1999, \prl, 82, 2832,
  \dodoi{10.1103/PhysRevLett.82.2832}

\bibitem[{{Shu} {et~al.}(2017){Shu}, {Brownstein}, {Bolton}, {Koopmans},
  {Treu}, {Montero-Dorta}, {Auger}, {Czoske}, {Gavazzi}, {Marshall}, \&
  {Moustakas}}]{shu2017slacs}
{Shu}, Y., {Brownstein}, J.~R., {Bolton}, A.~S., {et~al.} 2017, \apj, 851, 48,
  \dodoi{10.3847/1538-4357/aa9794}

\bibitem[{{Tulin} \& {Yu}(2018)}]{tulin2018dark}
{Tulin}, S., \& {Yu}, H.-B. 2018, \physrep, 730, 1,
  \dodoi{10.1016/j.physrep.2017.11.004}

\bibitem[{{Vegetti} {et~al.}(2018){Vegetti}, {Despali}, {Lovell}, \&
  {Enzi}}]{vegetti2018constraining}
{Vegetti}, S., {Despali}, G., {Lovell}, M.~R., \& {Enzi}, W. 2018, \mnras, 481,
  3661, \dodoi{10.1093/mnras/sty2393}

\bibitem[{{Vegetti} \& {Koopmans}(2009)}]{vegetti2009bayesian}
{Vegetti}, S., \& {Koopmans}, L.~V.~E. 2009, \mnras, 392, 945,
  \dodoi{10.1111/j.1365-2966.2008.14005.x}

\bibitem[{{Vegetti} {et~al.}(2010){Vegetti}, {Koopmans}, {Bolton}, {Treu}, \&
  {Gavazzi}}]{vegetti2010detection}
{Vegetti}, S., {Koopmans}, L.~V.~E., {Bolton}, A., {Treu}, T., \& {Gavazzi}, R.
  2010, \mnras, 408, 1969, \dodoi{10.1111/j.1365-2966.2010.16865.x}

\bibitem[{{Vegetti} {et~al.}(2012){Vegetti}, {Lagattuta}, {McKean}, {Auger},
  {Fassnacht}, \& {Koopmans}}]{vegetti2012gravitational}
{Vegetti}, S., {Lagattuta}, D.~J., {McKean}, J.~P., {et~al.} 2012, \nat, 481,
  341, \dodoi{10.1038/nature10669}

\bibitem[{{Viel} {et~al.}(2005){Viel}, {Lesgourgues}, {Haehnelt}, {Matarrese},
  \& {Riotto}}]{viel2005}
{Viel}, M., {Lesgourgues}, J., {Haehnelt}, M.~G., {Matarrese}, S., \& {Riotto},
  A. 2005, \prd, 71, 063534, \dodoi{10.1103/PhysRevD.71.063534}

\bibitem[{{Vogelsberger} {et~al.}(2019){Vogelsberger}, {Zavala}, {Schutz}, \&
  {Slatyer}}]{vogelsberger2019}
{Vogelsberger}, M., {Zavala}, J., {Schutz}, K., \& {Slatyer}, T.~R. 2019,
  \mnras, 484, 5437, \dodoi{10.1093/mnras/stz340}

\bibitem[{{Wagner-Carena} {et~al.}(2023){Wagner-Carena}, {Aalbers}, {Birrer},
  {Nadler}, {Darragh-Ford}, {Marshall}, \& {Wechsler}}]{wagnercarena2022fidm}
{Wagner-Carena}, S., {Aalbers}, J., {Birrer}, S., {et~al.} 2023, \apj, 942, 75,
  \dodoi{10.3847/1538-4357/aca525}

\bibitem[{{Wagner-Carena} {et~al.}(2021){Wagner-Carena}, {Park}, {Birrer},
  {Marshall}, {Roodman}, {Wechsler}, \& {LSST Dark Energy Science
  Collaboration}}]{wagner2021hierarchical}
{Wagner-Carena}, S., {Park}, J.~W., {Birrer}, S., {et~al.} 2021, \apj, 909,
  187, \dodoi{10.3847/1538-4357/abdf59}

\bibitem[{{White}(2001)}]{white2001mass}
{White}, M. 2001, \aap, 367, 27, \dodoi{10.1051/0004-6361:20000357}

\bibitem[{{White} \& {Rees}(1978)}]{white1978core}
{White}, S.~D.~M., \& {Rees}, M.~J. 1978, \mnras, 183, 341,
  \dodoi{10.1093/mnras/183.3.341}

\bibitem[{{Zhang} {et~al.}(2024){Zhang}, {{\c{S}}eng{\"u}l}, \&
  {Dvorkin}}]{zhang2024densityslope}
{Zhang}, G., {{\c{S}}eng{\"u}l}, A.~{\c{C}}., \& {Dvorkin}, C. 2024, \mnras,
  527, 4183, \dodoi{10.1093/mnras/stad3521}

\end{thebibliography}

\appendix

\begin{table*}
    \centering
    \renewcommand{\arraystretch}{1.3}
    \begin{tabularx}{\textwidth}{@{} X X @{}}
    \toprule
    \textbf{Component} & \textbf{Prior / Training  Distribution -- $p(\theta|\Omega_0)$} \\
    \midrule
    \raggedright \textbf{Main Deflector} &  \\
    
    \textcolor{purple}{x-coordinate lens center ($\arcsec$)} & \textcolor{purple}{$x_\text{lens} \sim \mathcal{N}(\mu: 0, \sigma: 0.16)$} \\
    
    \textcolor{purple}{y-coordinate lens center ($\arcsec$)} & \textcolor{purple}{$y_\text{lens} \sim \mathcal{N}(\mu: 0,\sigma: 0.16)$} \\
    
    \textcolor{purple}{Einstein Radius ($\arcsec$)} & \textcolor{purple}{$\theta_E \sim \mathcal{N}(\mu: 1.1, \sigma: 0.15)$} \\
    
    \textcolor{purple}{Power-law slope} & \textcolor{purple}{$\gamma_\text{lens} \sim \mathcal{N}(\mu: 2.0,\sigma: 0.1)$} \\
    
    \raggedright \textcolor{purple}{x-direction ellipticity eccentricity} & \textcolor{purple}{$e_1 \sim \mathcal{N}(\mu: 0,\sigma: 0.1)$} \\
    
    \raggedright \textcolor{purple}{xy-direction ellipticity eccentricity} & \textcolor{purple}{$e_2 \sim \mathcal{N}(\mu: 0,\sigma: 0.1)$} \\
    
    Main halo critical mass $(M_\odot)$ & $m_\text{host} = 10^{13}$ \\
    
    Main halo redshift & $z_{\text{lens}} = 0.5$ \\
    
    \textcolor{purple}{x-direction shear} & \textcolor{purple}{$\gamma_1 \sim \mathcal{N}(\mu: 0,\sigma: 0.05)$} \\
    
    \textcolor{purple}{xy-direction shear} & \textcolor{purple}{$\gamma_2 \sim \mathcal{N}(\mu: 0,\sigma: 0.05)$} \\
    
    \midrule
    \textbf{Mass--concentration} &  \\
    \raggedright Concentration normalization & $c_0 = \mathcal{N}(\mu: 16, \sigma: 2)$ \\
    
    \raggedright Redshift power-law slope & $\zeta = \mathcal{N}(\mu: -0.3, \sigma: 0.1)$ \\
    
    \raggedright Peak height power-law slope & $\beta = \mathcal{N}(\mu: 0.55, \sigma: 0.3)$ \\
    
    \raggedright mass--concentration power-law pivot mass ($M_\odot$) & $m_{\text{pivot,conc}} = 10^{8}$ \\
    
    Concentration dex scatter & $\sigma_{\text{conc}} = \mathcal{N}(\mu: 0.1, \sigma: 0.06)$ \\
    
    \midrule
    \textbf{Cosmology} &  \\
    Cosmology Assumption & $\Lambda$CDM from Planck 2018 \\
    
    \midrule
    \textbf{Subhalos} &  \\
    \raggedright Subhalo mass function power-law index & $\gamma_{\text{sub}} \sim \text{Unif}(-2.02,-1.92)$ \\
    
    \textcolor{purple}{\raggedright Subhalo mass function normalization $(\text{kpc}^{-2})$} & 
    \textcolor{purple}{$\Sigma_{\text{sub}} \sim \mathcal{N}(\mu: 2 \times 10^{-3},\sigma: 1.1 \times 10^{-3})$} \\
    
    \raggedright Subhalo power-law pivot mass ($M_\odot$) &
    $m_{\text{pivot,sub}} = 10^{10}$ \\
    
    \raggedright Subhalo mass function minimum mass ($M_\odot$) &
    $m_{\text{min,sub}} = 10^{7}$ \\
    
    \raggedright Subhalo mass function maximum mass ($M_\odot$) &
    $m_{\text{max,sub}} = 10^{10}$ \\
    
    \raggedright Subhalo truncation pivot mass ($M_\odot$) &
    $m_{\text{pivot,trunc}} = 10^{7}$ \\
    
    \raggedright Subhalo truncation pivot radius (kpc) &
    $r_{\text{pivot,trunc}} = 50$ \\
    
    \midrule
    \raggedright \textbf{Source: COSMOS catalog}  &  \\
    
    Source redshift & $ z_{\text{source}} = 1.5$ \\
    
    Maximum catalog redshift & $ z_{\text{catalog,max}} = 1.0$ \\
    
    Faintest catalog apparent magnitude & $ \text{mag}_{\text{faint}} = 20$ \\
    
    Minimum source size (pixels) & $\text{size}_\text{min,pix} = 50$ \\
    
    Minimum half-light radius (pixels) & $r_{1/2} = 10$ \\
    
    Source rotation angle & $\phi_\text{source} \sim \text{Unif}(0,2\pi)$ \\
    
    Source amplitude multiplier & $a_\text{source} \sim \text{Unif}(0.5,2.0)$ \\
    
    \textcolor{purple}{x-coordinate source center ($\arcsec$)} & \textcolor{purple}{$x_\text{source} \sim \mathcal{N}(\mu: 0, \sigma: 0.16)$} \\
    
    \textcolor{purple}{y-coordinate source center ($\arcsec$)} & \textcolor{purple}{$y_\text{source} \sim \mathcal{N}(\mu: 0,\sigma: 0.16)$} \\
    
    Number of galaxy images & Training: 2,163 / Testing: 99 \\
    
    \bottomrule
    \end{tabularx}
    \caption{Distribution of simulation parameters in the training set.}
    \label{tab:dist}
    \tablecomments{The subsets of galaxy images used for the training is disjoint from the galaxy images used for all the plots and metrics presented throughout the work. In this table, $\mathcal{N}$ is the normal distribution and  $\text{Unif}$ is the uniform distribution. Parameters highlighted in purple are inferred by the NPE / SNPE model.}
\end{table*}

\section{Jax Implementation Choices}\label{app:jax_implementation}

While the use of \textsc{jax} enables substantial performance improvements, it also requires a few implementation compromises. First, \textsc{paltax} is designed around a very rigid assumption about the lensing geometry: the mass profiles are assumed to come from line-of-sight halos, subhalos, the main deflector, and then line-of-sight halos in that order. This avoids a limitation with branching. In \textsc{jax}, branching generally leads to both computational branches being executed. Therefore, if you allow any mass profile to be used, as is done in lensing packages like \textsc{lenstronomy}, every deflector will execute a calculation over every mass profile in your library. Instead, by specifying the order of deflector categories a priori, only the profiles appropriate for that category of deflector will be evaluated. The \textsc{paltax} code can be modified to work on alternate configurations, but it is not as trivial as modifying a configuration file. Second, a number of the cosmological calculations relevant to the lensing code are repeated for each simulation. In order to limit the overhead of the cosmology library, these values are pre-computed and stored. When a query doesn't match one of the pre-computed values, a linear interpolation is used. The reference cosmology code, \textsc{colossus}, uses similar lookup table tricks to speed up its own calculations. Finally, \textsc{jax} compilation deals poorly with variable length arrays and loops. To ensure optimal performance, the maximum number of line-of-sight halos\footnote{For line-of-sight halos, this maximum applies to each redshift bin.}, subhalos, and samples used for rejection sampling must be specified ahead of time.

\section{Simulation Parameter Distributions}\label{app:sim_param_distributions}

In Table \ref{tab:dist} we present the distribution of parameters used to draw training examples for the model. Note that SNPE explicitly modifies the distribution of the parameters inferred by the model. These are highlighted in purple in the table.

\section{Sequential Training Overfitting}\label{app:seq_train_overfit}

\begin{figure*}
    \centering
	\includegraphics[width=0.75\textwidth]{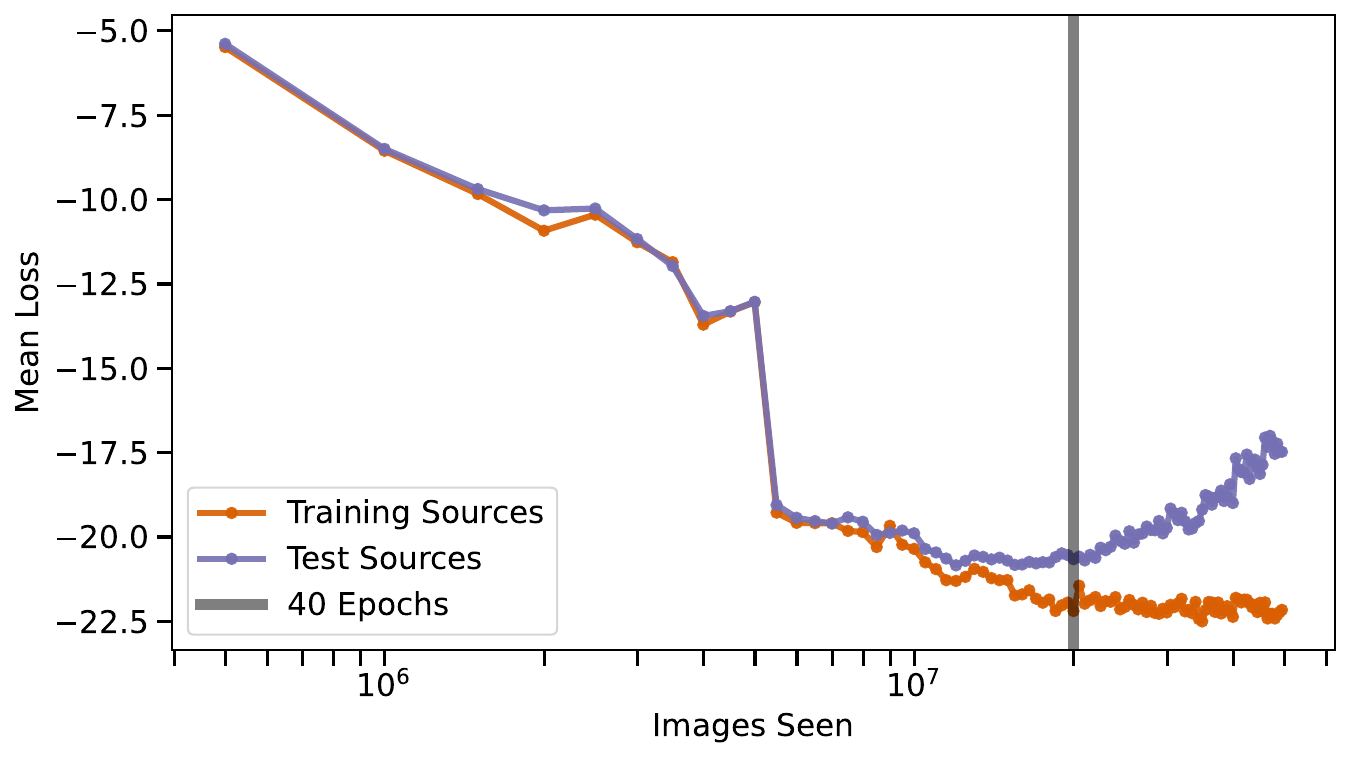}
    \caption{A comparison of the SNPE loss over all parameters for thirty batches of 32 images, with each batch tightly centered on one of the thirty mock lenses. Two image generation schemes are compared: one that uses the source images seen during training (orange) and one that uses the source images held out for testing (purple). The forty epoch stopping point used in Section \ref{sec:seq_inf_results} is shaded in black.}
    \label{fig:source_loss_comp}
\end{figure*}
\begin{figure*}
    \centering
	\includegraphics[width=0.67\textwidth]{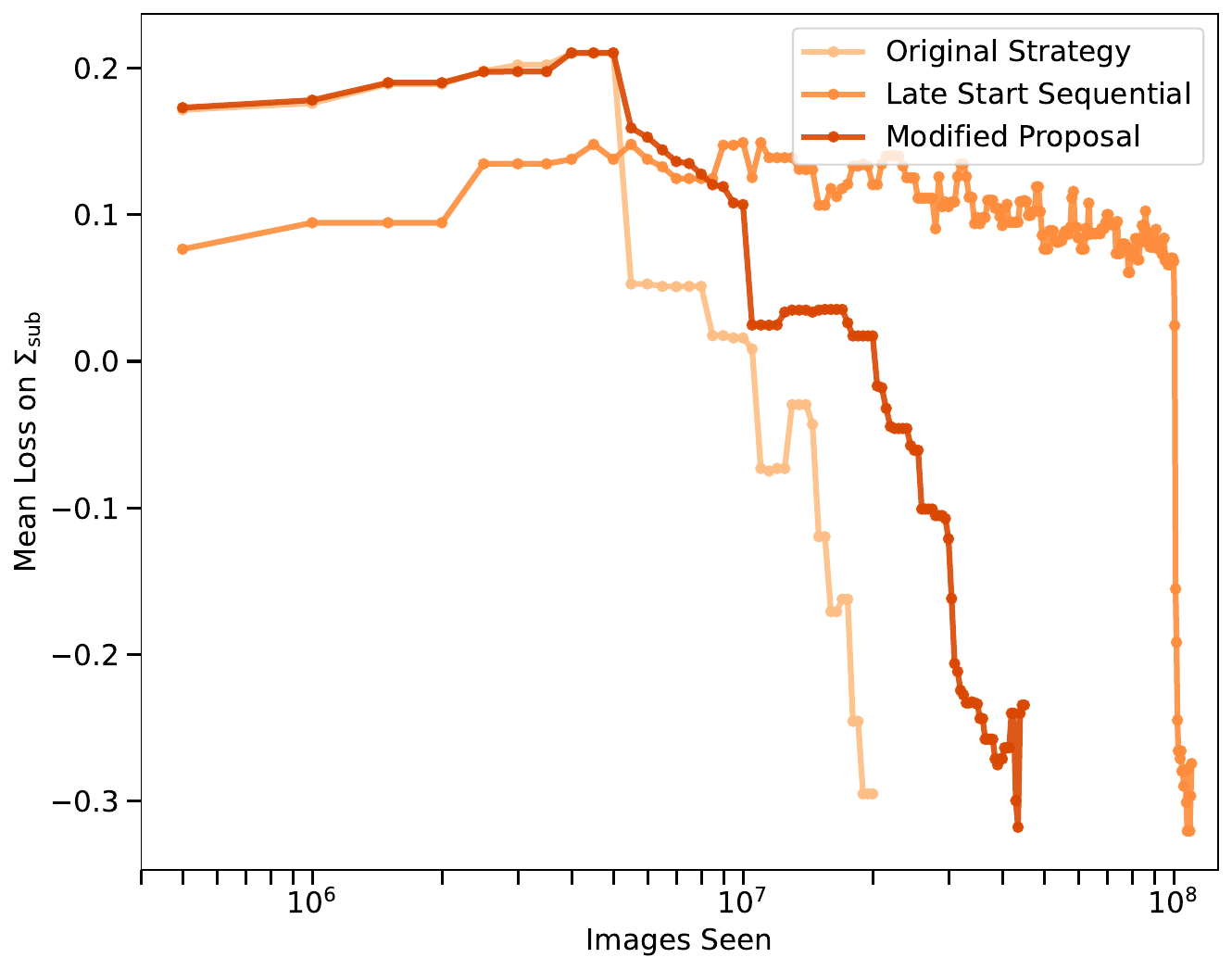}
    \caption{A comparison of the mean loss on $\Sigma_\text{sub}$ on 6 mock observations for three different sequential strategies: 10 epochs of broad training with the original proposal strategy (Equation \ref{eq:orig_prop_dist}, light orange), 200 epochs of broad training with the original proposal strategy (orange), and 10 epochs of broad training with a weighted mean proposal strategy (Equation \ref{eq:new_prop_dist}, dark orange). Each sequential strategy is stopped when it begins exhibiting overfitting on the training source distribution (see Appendix \ref{app:seq_train_overfit}).}
    \label{fig:loss_seq_strat_comp}
\end{figure*}

In Section \ref{sec:seq_inf_results} we present our SNPE results up to 40 epochs of training. The choice to cut our SNPE method at this point is determined by the shortcomings of our source model. Unlike the parameters of the main deflector which are continuous, our source model is drawn from a set of 2,163 cosmos images. The amplitude, position, and angle of these sources are randomly sampled, but the fundamental morphology is set by the discrete samples in our catalog. To ensure that we do not overstate the capabilities of our model, we hold out 99 sources for all of the metrics and plots we present throughout the paper. The 30 mock observations are also generated using these held-out sources.

We would like to probe the extent to which our sequential model has overfit to the specific sources in our training catalog. We run each of our thirty models on thirty batches of 32 test images. Each batch is generated with lensing parameters tightly clustered around the true value of our mock observations. We compare how the model performs if these lenses are drawn using the training source catalog or the test source catalog. Essentially, we generate lensing images meant to closely mimic the mock observations up to the underlying source model. In Figure \ref{fig:source_loss_comp} we compare the loss under our two different source distribution assumptions. Until around thirty epochs, the performance of our sequential models is robust to shifting the underlying source distribution, but between forty and thirty epochs, the two loss curves begin to diverge. Beyond the forty epoch limit, our loss on the test sources begins to grow despite improvements in the loss on our training sources. Since these two sets of lensing images are identical up to the source distribution, we conclude that forty epochs is the appropriate early stopping choice to avoid overfitting to the training source catalog.

If we want to push our sequential models beyond the current performance, we believe we will need a larger, more diverse source catalog. Possible improvements to our approach to the source distribution are discussed in Section \ref{sec:discussion}. 

\begin{figure*}
    \centering
	\includegraphics[width=0.8\textwidth]{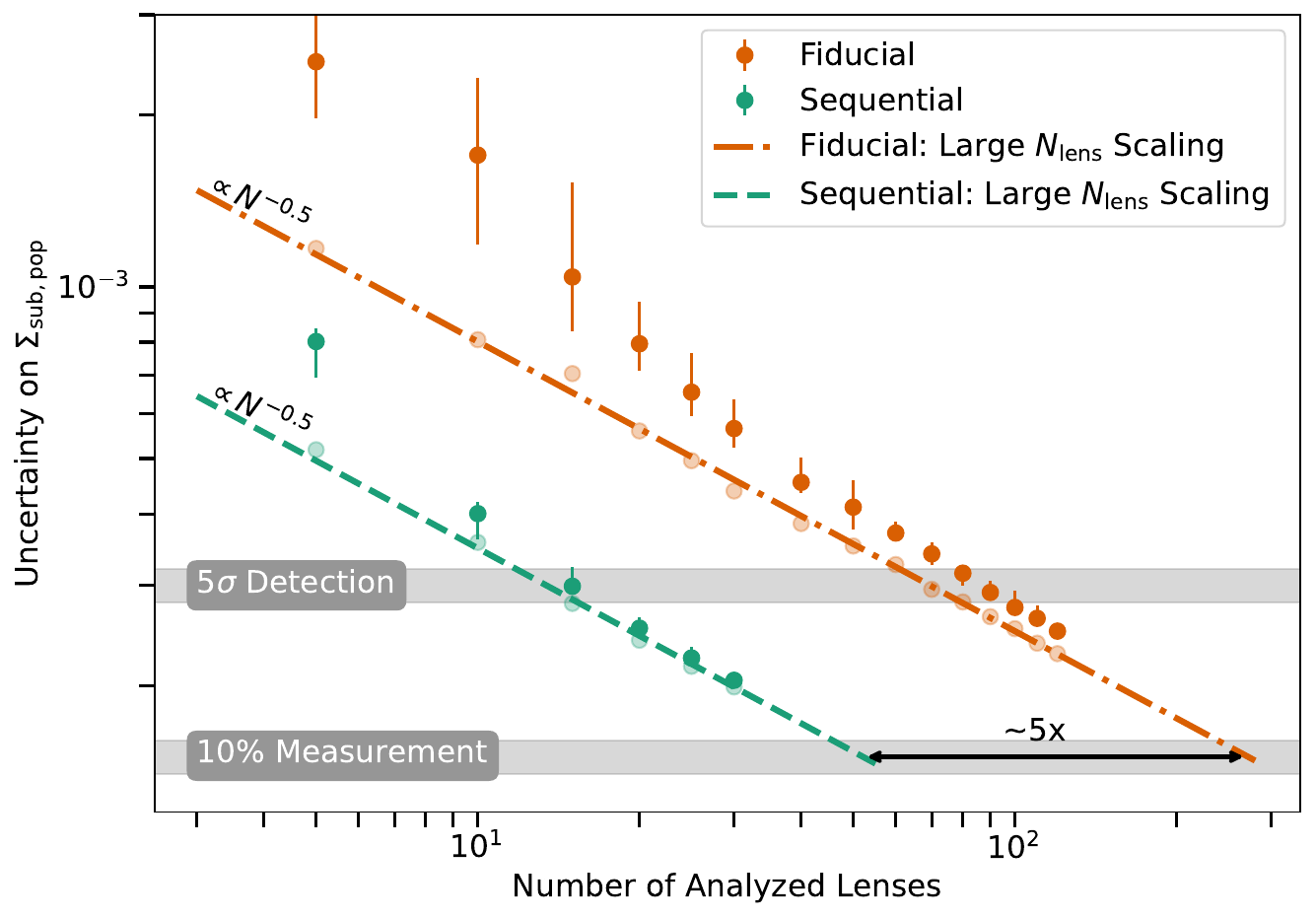}
    \caption{The same one-sigma uncertainty on the $\Sigma_\text{sub,pop}$ population constraint as a function of the number of lenses in the analysis presented in Figure \ref{fig:scaling_hierarchical}. The plot now includes the median hierarchical constraints with fixed $\Sigma_{\text{sub,pop},\sigma}$ used to fit the power-law relationship.}
    \label{fig:scaling_hierarchical_detailed}
\end{figure*}

\section{Sequential Proposal Distribution}\label{app:seq_proposal}

For the sequential inference results presented in the paper, we used a simple rule for generating a new proposal distribution:
\begin{align}\label{eq:orig_prop_dist}
    p_{j}(\theta | \Omega_i) = q_{\phi,i-1}(\theta|x_j, \Omega_{0}),
\end{align}
where the $j$ index corresponds to the observation and the $i$ index corresponds to the round of sequential inference. Since the model generated by the SNPE-C approach always approximates $p(\theta|x,\Omega_0)$, the only strict requirement for our sequential approach is that the normalizing factor in Equation \ref{eq:SNPE-C} be analytic\footnote{See \cite{greenberg2019automatic} for how to circumvent this requirement.}. Similarly, we also chose to start our sequential inference after 10 epochs of training on the broad distribution. Since training on the broad distribution doesn't scale with the number of lenses being analyzed, it may be beneficial to train on the broad distribution for longer before beginning the sequential training.

Here, we show the results of modifying those two choices. First, we use a new scheme for selecting the proposal distribution:
\begin{align}\label{eq:new_prop_dist}
    p_{j}(\theta | \Omega_i) = (1-\lambda_\text{prop}) \times p_{j}(\theta | \Omega_{i-1}) + \lambda_\text{prop} \times q_{\phi,i-1}(\theta|x_j, \Omega_{0}).
\end{align}
In this scheme, each new proposal is a weighted average of the previous proposal and the current posterior estimate. Since our proposals and posteriors are Gaussian, our new proposal distributions are a mixture of Gaussian distributions. Notably, the original, broad distribution is carried forward between proposals with an exponential decay of $(1-\lambda_\text{prop})^{i}$. We also experiment with keeping the proposal distribution from Equation \ref{eq:orig_prop_dist} but beginning our sequential training after 200 epochs of training on the broad distribution.

The mean loss on $\Sigma_\text{sub}$ is shown in Figure \ref{fig:loss_seq_strat_comp}. Note that this analysis is run on only 6 of our 30 mock observations. The two new sequential approaches are stopped when the model begins to overfit on the training source distribution (see Appendix \ref{app:seq_train_overfit}). For the modified sequential proposal strategy we set $\lambda_\text{prop} = 0.5$. All three choices lead to equivalent performance on $\Sigma_\text{sub}$, with the approach used in the main body requiring the fewest number of total training steps. Because the performance limit appears to be driven by the training source distribution, we reserve a more robust analysis of the hyperparameters of the sequential training for future work.

\begin{figure*}
    \centering
    \includegraphics[width=0.45\textwidth]{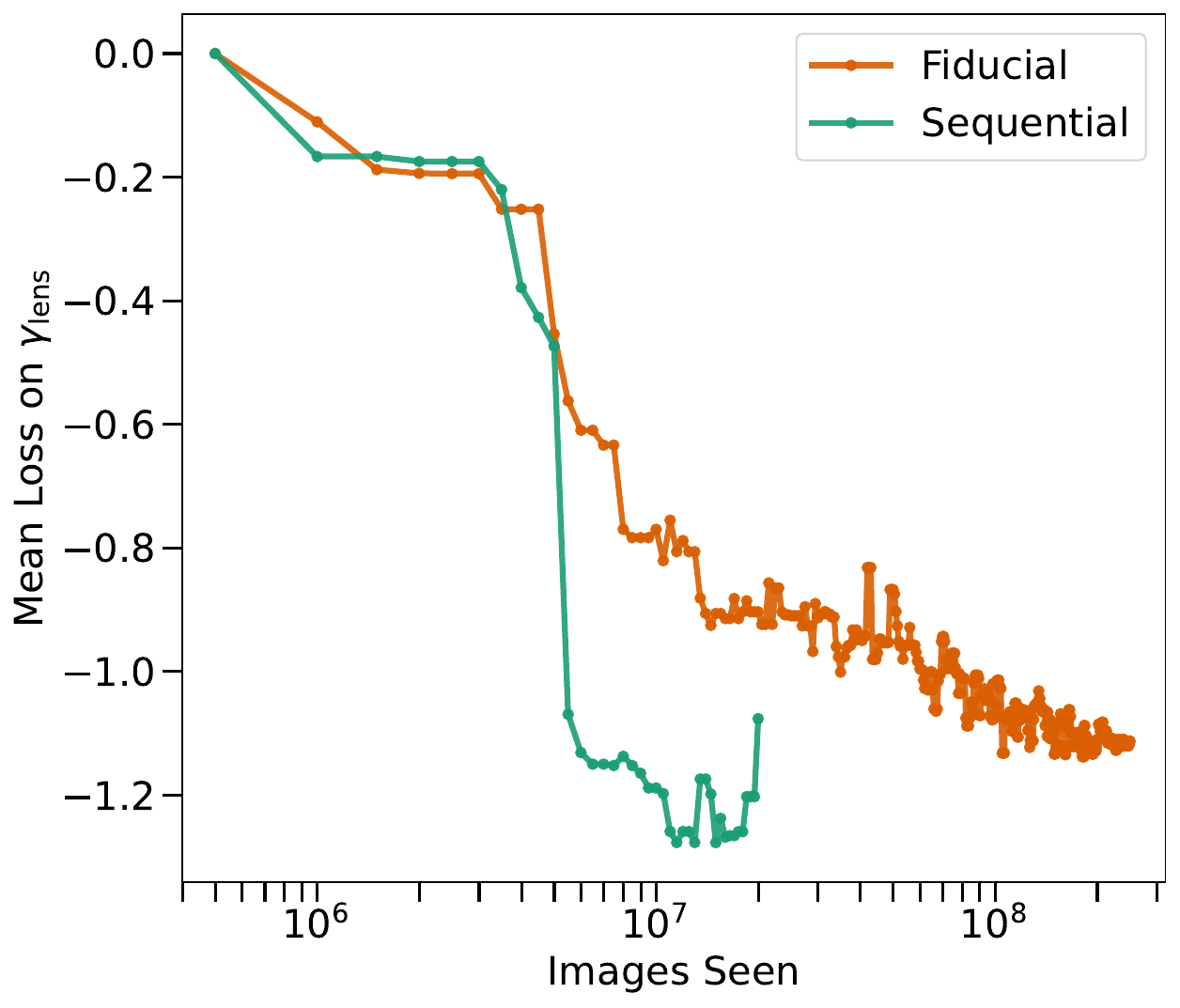}
    \includegraphics[width=0.45\textwidth]{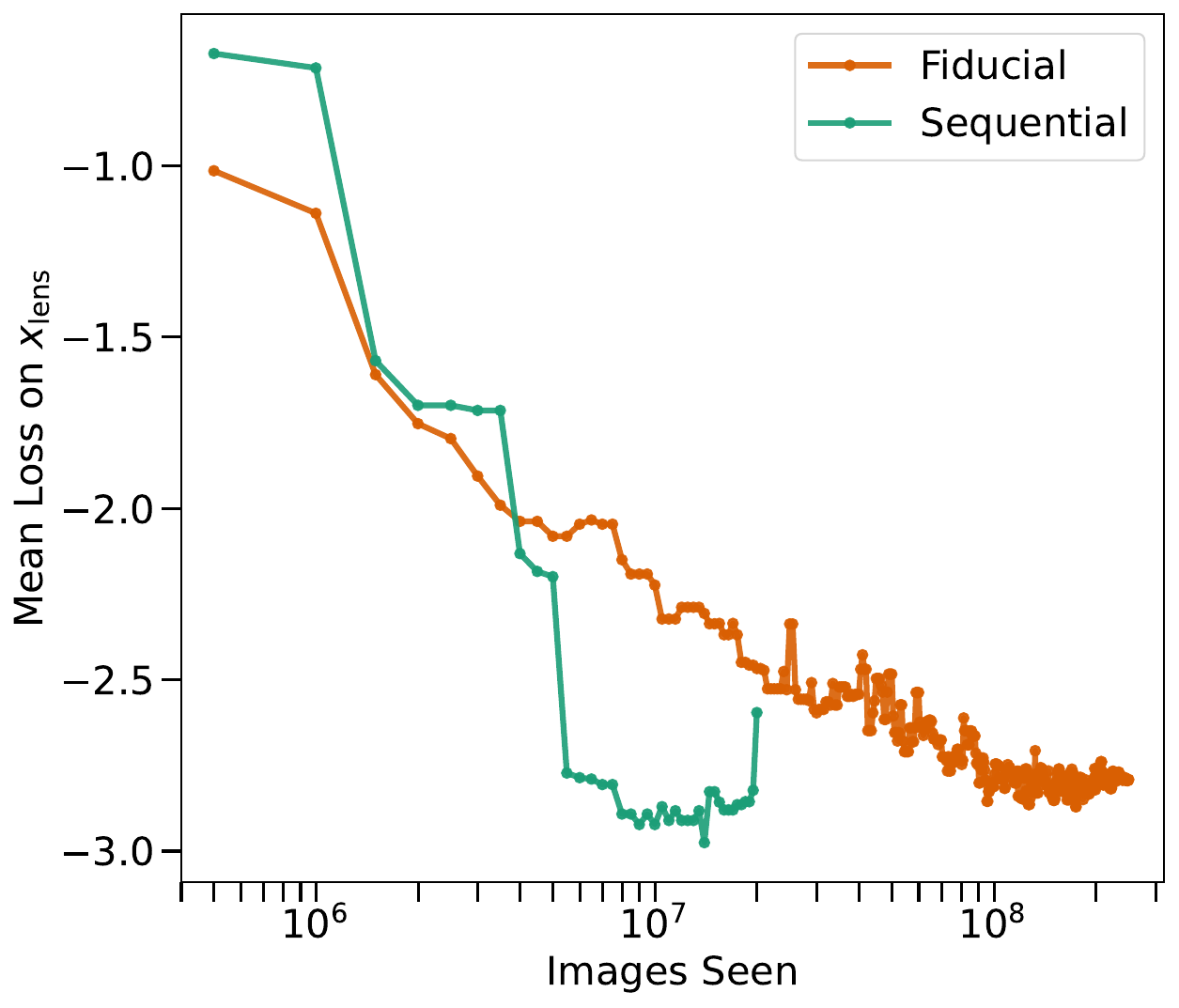}
    \includegraphics[width=0.45\textwidth]{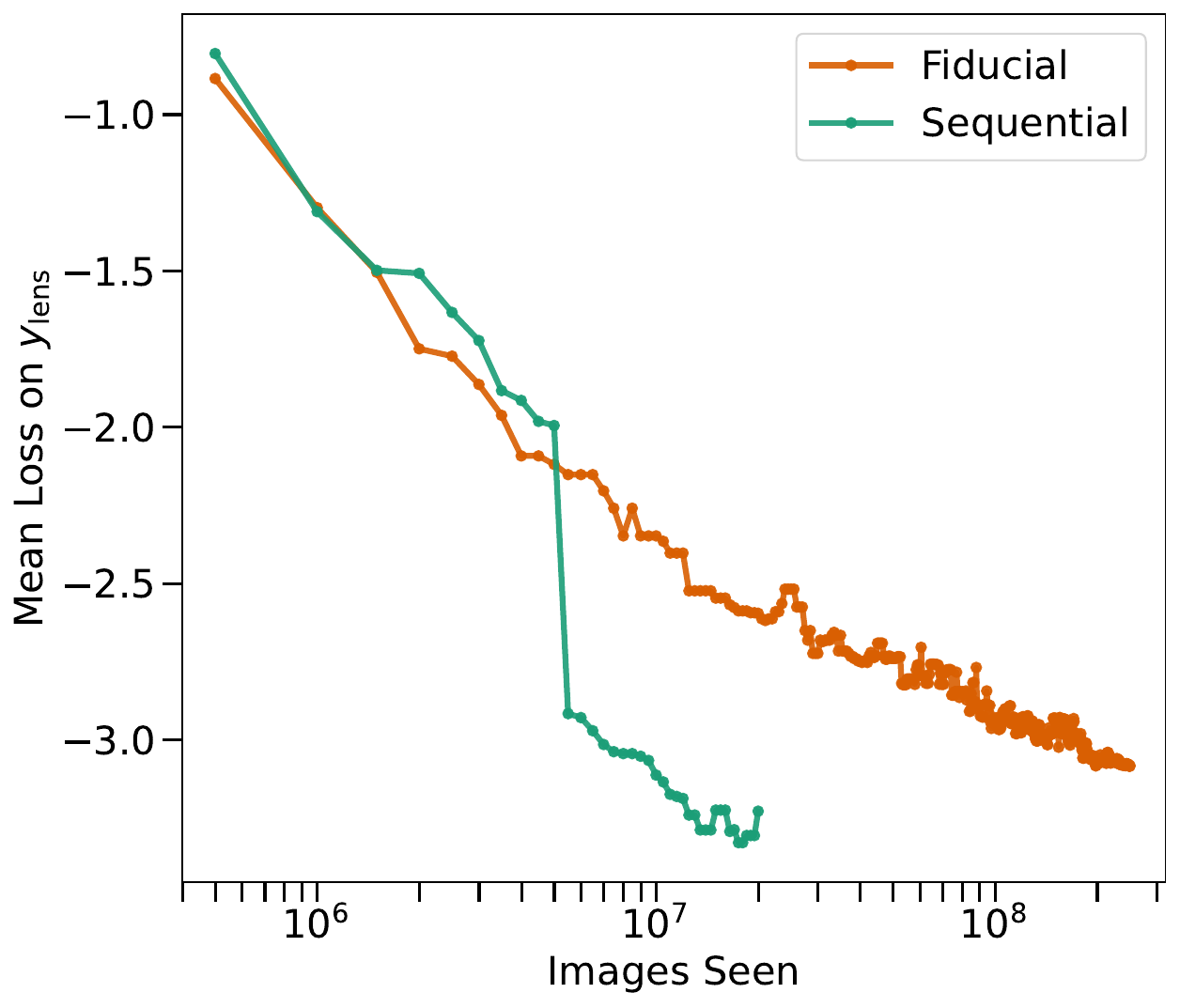}
    \includegraphics[width=0.45\textwidth]{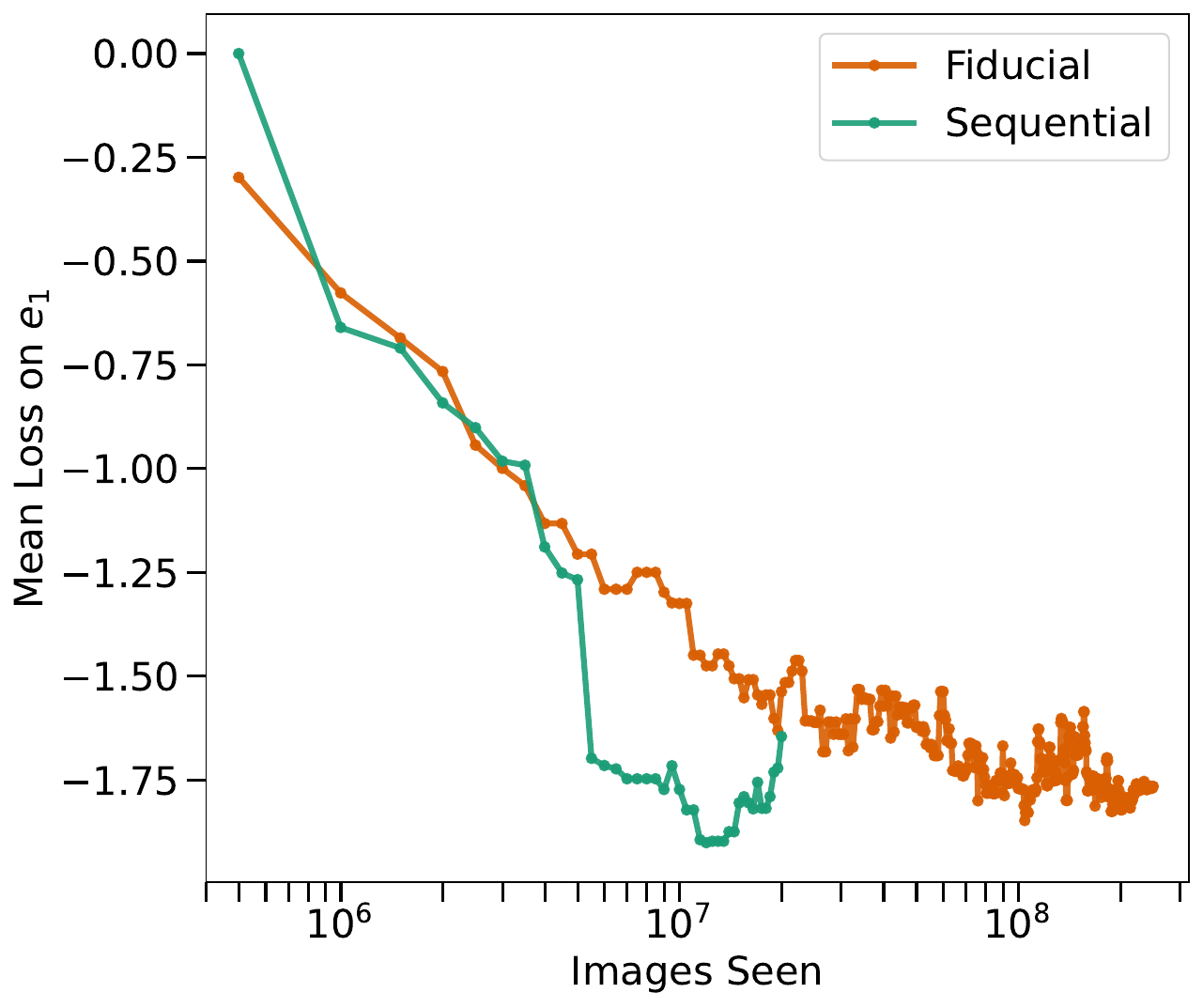}
    \caption{A comparison of the mean loss on the power-law slope, the x-coordinate lens center, the y-coordinate lens center, and the x-direction ellipticity shear -- $\gamma_\text{lens}, x_\text{lens}, y_\text{lens},$ and $e_1$ -- on 30 mock observations for two different methodologies: the fiducial, NPE approach (orange) and the sequential, SNPE approach (green).}
    \label{fig:seq_add_params_one}
\end{figure*}

\begin{figure*}
    \centering
    \includegraphics[width=0.45\textwidth]{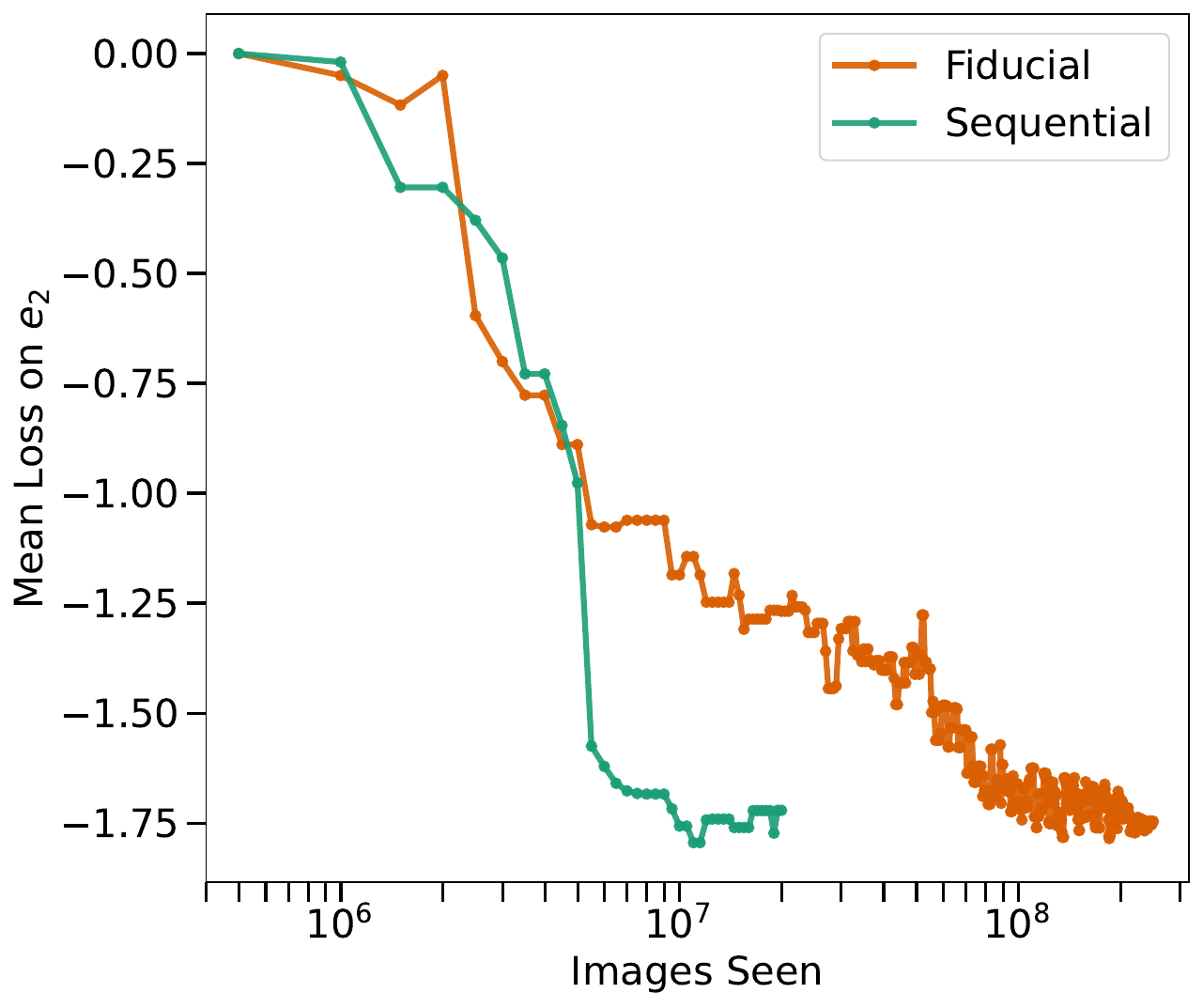}
    \includegraphics[width=0.45\textwidth]{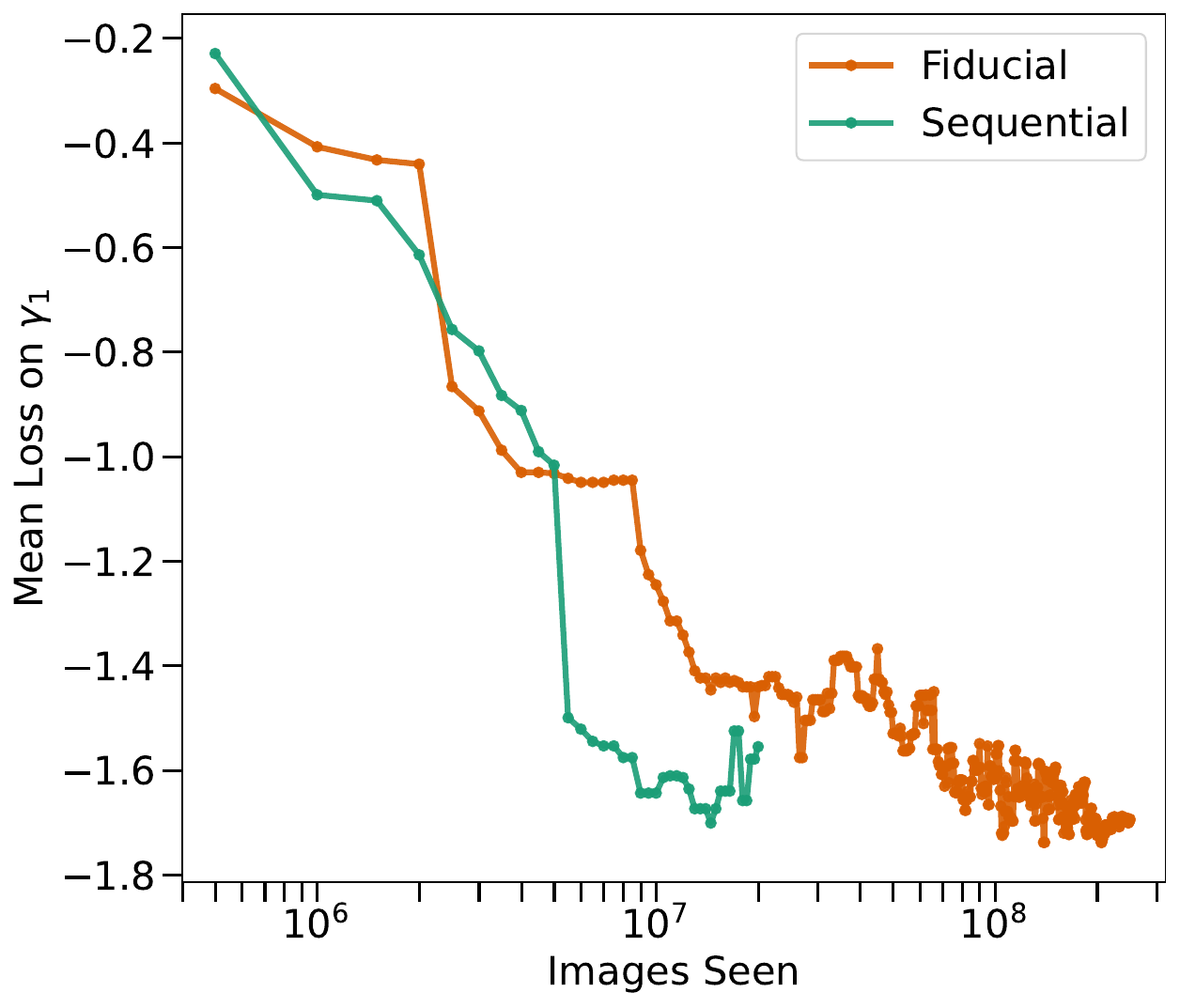}
    \includegraphics[width=0.45\textwidth]{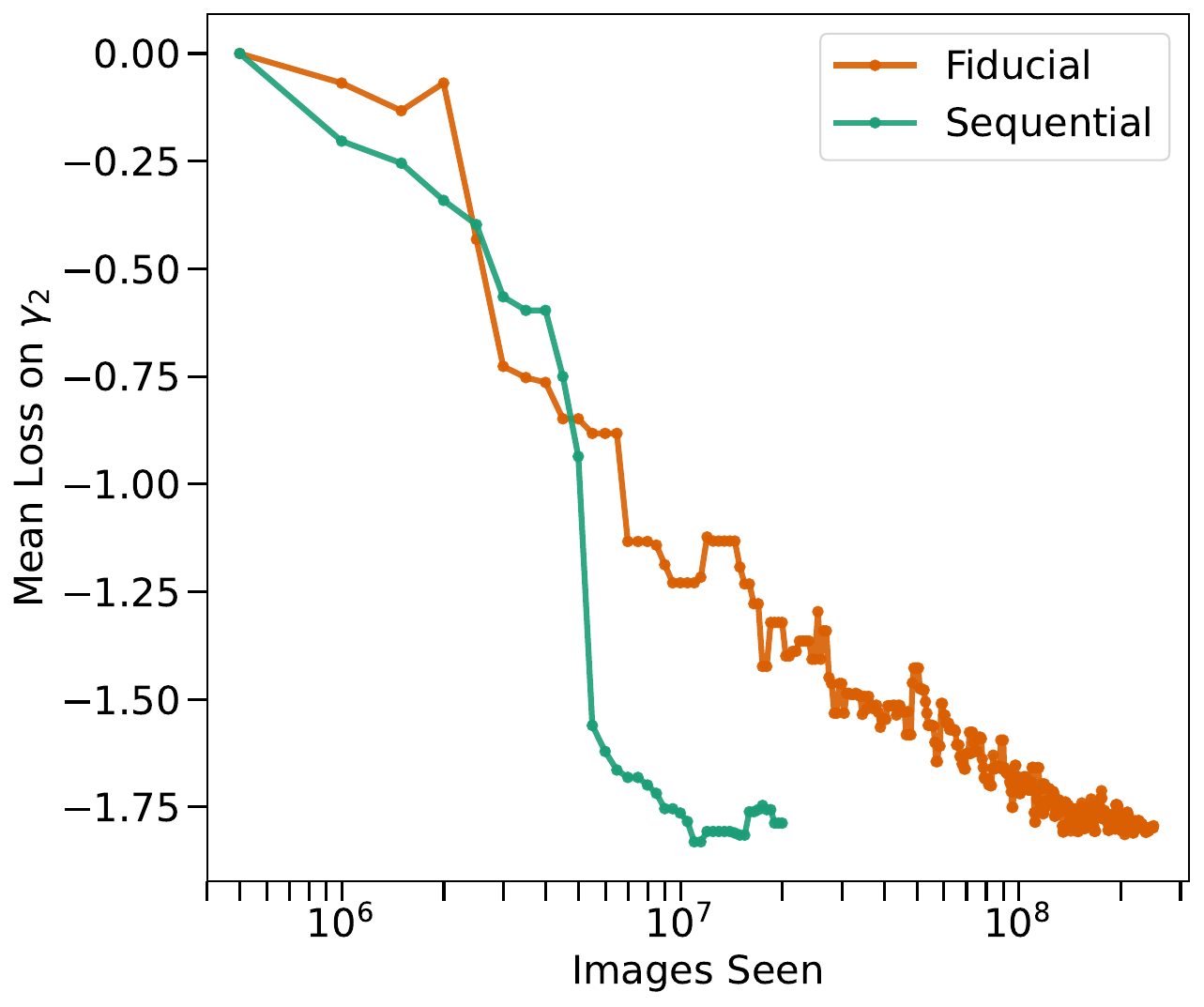}
    \includegraphics[width=0.45\textwidth]{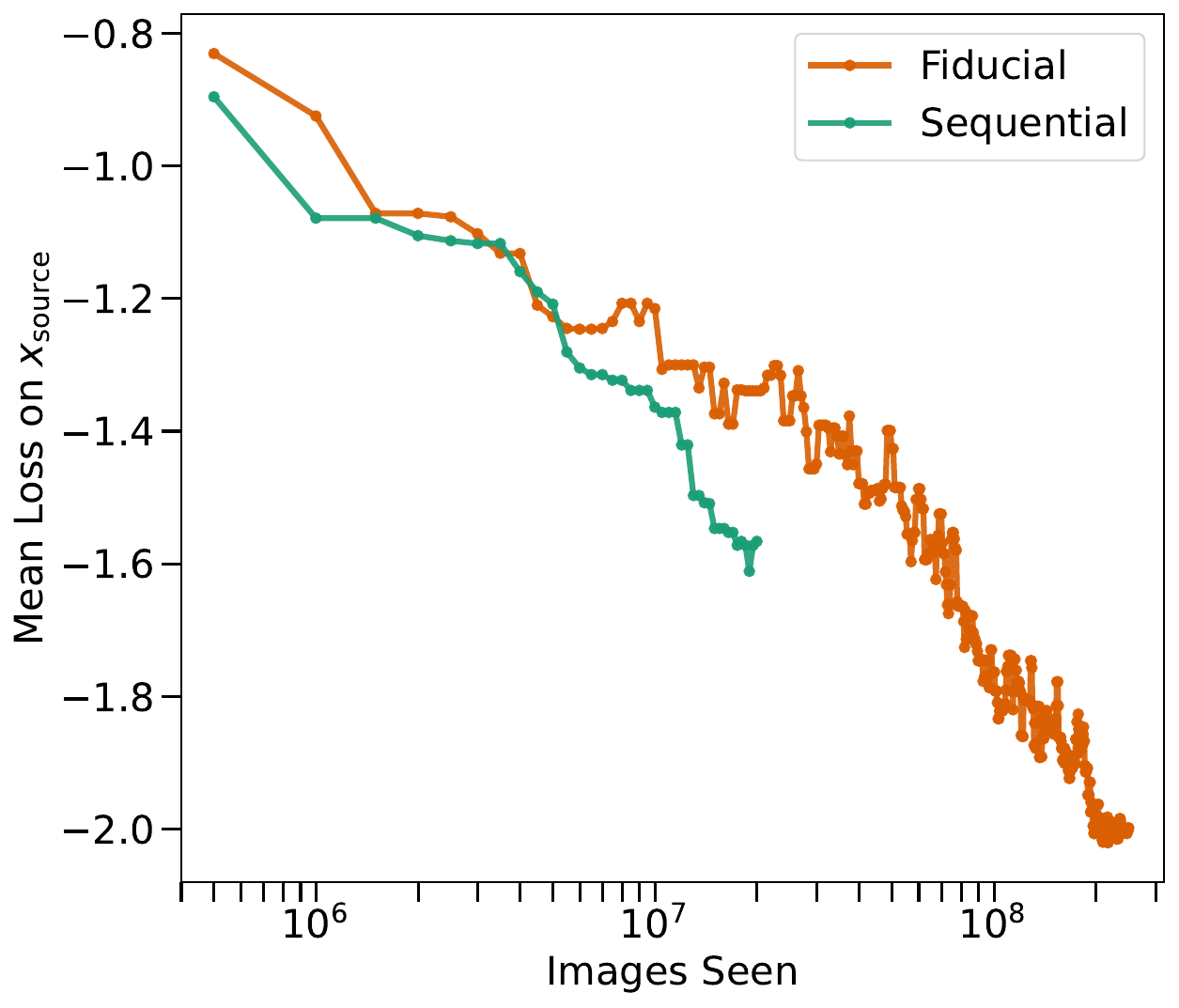}
    \includegraphics[width=0.45\textwidth]{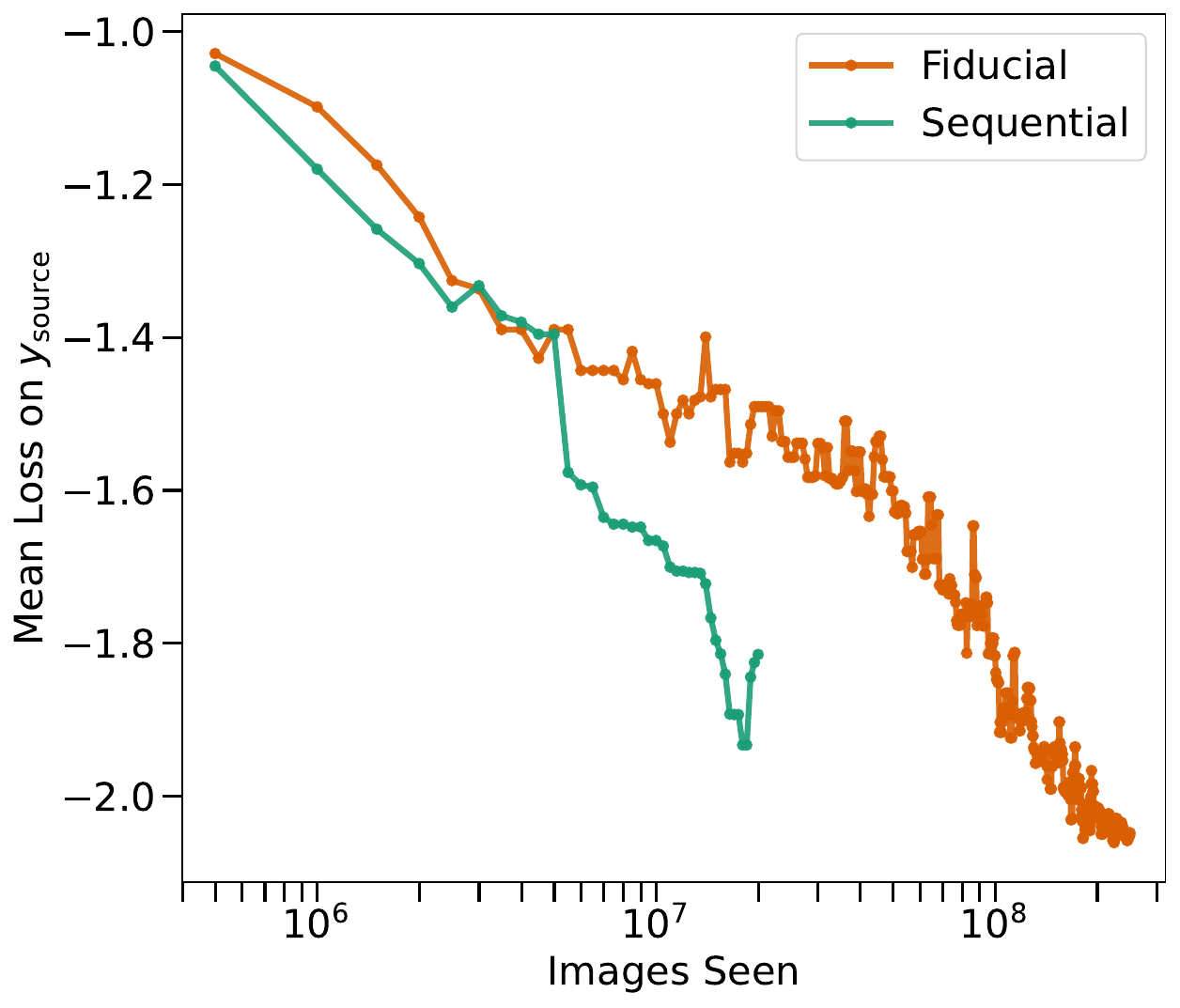}
    \caption{A comparison of the mean loss on the xy-direction ellipticity eccentricity, the x-direction shear, the xy-direction shear, the x-coordinate source center, and the y-coordinate source center -- $e_2, \gamma_1, \gamma_2, x_\text{source},$ and $ y_\text{source}$ -- on 30 mock observations for two different methodologies: the fiducial, NPE approach (orange) and the sequential, SNPE approach (green).}
    \label{fig:seq_add_params_two}
\end{figure*}

\section{Large Lens Scaling of Hierarchical Constraints}\label{app:large_n_hier_const}

In Figure \ref{fig:scaling_hierarchical} we include two lines that capture the large $N_\text{lens}$ scaling of the one-sigma uncertainties on $\Sigma_\text{sub,pop}$. These lines are fit to a hierarchical analysis where $\Sigma_{\text{sub,pop},\sigma}$ is treated as known. To see why, consider the simpler case of $N$ measurements each with identical Gaussian error $\sigma_\text{meas}$ drawn from a Gaussian population with standard deviation $\sigma_{pop}$. The variance of the mean estimator, $\hat{\mu}_\text{pop}$, is given by:
\begin{align}\label{eq:app_mu_pop}
    \text{var}[\hat{\mu}_\text{pop}] = \frac{\sigma_\text{meas}^2 + \sigma_\text{pop}^2}{N}.
\end{align}
If we know $\sigma_\text{pop}$, then the variance on our mean estimator scales like $\frac{1}{N}$. However, our hierarchical analysis is also inferring $\sigma_\text{pop}$. When the maximum inferred $\sigma_\text{pop}$, $\sigma_\text{pop,max}$, has the property that $\sigma_\text{pop,max}<<\sigma_\text{meas}$, then the numerator in Equation \ref{eq:app_mu_pop} is essentially unchanged by a tighter constraint on $\sigma_\text{pop}$. However, if our posterior region for $\sigma_\text{pop}$ transitions from $\sigma_\text{pop,max} \approx \sigma_\text{meas}$ to $\sigma_\text{pop,max} << \sigma_\text{meas}$ this induces an additional $N$ scaling for $\text{var}[\hat{\mu}_\text{pop}]$. 

This is exactly the behavior of our hierarchical inference. First, the true population scatter, $\Sigma_{\text{sub,pop},\sigma}$, is smaller than the uncertainties our model outputs on $\Sigma_\text{sub}$ for an individual lens. Second, as we see in Figure \ref{fig:hier_inf}, large $\Sigma_{\text{sub,pop},\sigma}$ values correlate with larger uncertainties on $\Sigma_{\text{sub,pop}}$. As our constraints on $\Sigma_{\text{sub,pop},\sigma}$ become tighter, this correlation between $\Sigma_{\text{sub,pop},\sigma}$ and the uncertainty in $\Sigma_{\text{sub,pop}}$ disappears. This agrees with our expectations from Equation \ref{eq:app_mu_pop}: for small $N$, where the numerator is decreasing with $N$, our scaling on the variance should be better than $\frac{1}{N}$. For large $N$, the numerator is dominated by $\sigma_\text{meas}$, so the scaling settles into $\frac{1}{N}$. Therefore, we expect an analysis with fixed $\Sigma_{\text{sub,pop},\sigma}$ to more accurately capture the large $N_\text{lens}$ scaling of the uncertainty on $\Sigma_{\text{sub,pop}}$.

In Figure \ref{fig:scaling_hierarchical_detailed} we recreate Figure \ref{fig:scaling_hierarchical} but include the uncertainties derived when $\Sigma_{\text{sub,pop},\sigma}$ is fixed. These partially translucent points were used to fit the power-law scaling. Note that the $\frac{1}{\sqrt{N}}$ of the scaling agrees with our expectation from Equation \ref{eq:app_mu_pop} but was a result of the fit.

\section{Sequential Comparison Additional Parameters}\label{app:seq_comp_add_params}

In Figure \ref{fig:seq_add_params_one} and Figure \ref{fig:seq_add_params_two} we compare the loss between the fiducial and sequential approaches for the parameters not presented in Section \ref{sec:seq_inf_results}.

\end{document}